\documentclass[prb,twocolumn,showpacs,preprintnumbers,amsmath,amssymb,floatfix]{revtex4}
\usepackage{graphicx}
\usepackage{float}
\usepackage{trfsigns}

\newcommand \be{\begin{eqnarray}}
\newcommand \ee{\end{eqnarray}}

\newcommand \ba{\begin{align}}
\newcommand \eea{\end{align}}

\newcommand {\p}[1]{\partial_{#1}}

\newcommand \V{\vec}

\begin{document}
           \csname @twocolumnfalse\endcsname
\title{Extended quasiparticle Pad\'e approximation for non-Fermi liquids}
\author{Klaus Morawetz$^{1,2}$
}
\affiliation{$^1$M\"unster University of Applied Sciences,
Stegerwaldstrasse 39, 48565 Steinfurt, Germany}
\affiliation{$^2$International Institute of Physics- UFRN,
Campus Universit\'ario Lagoa nova,
59078-970 Natal, Brazil
}

\begin{abstract}
The extended quasiparticle picture is adapted to non-Fermi systems by suggesting a Pad\'e approximation which interpolates between the known small scattering-rate expansion and the deviation from the Fermi energy. The first two energy-weighted sum rules are shown to be fulfilled independent of the interpolating function for any selfenergy. For various models of one-dimensional Fermions scattering with impurities the quality of the Pad\'e approximation for the spectral function is demonstrated and the reduced density matrix or momentum distribution is reproduced not possessing a jump at the Fermi energy. Though the two-fold expansion is necessary to realize the spectral function and reduced density, the extended quasiparticle approximation itself is sufficient for the description of transport properties due to cancellation of divergent terms under integration. The T-matrix approximation leads to the delay time as the time two particles spend in a correlated state. This contributes to the reduced density matrix and to an additional part in the conductivity which is presented at zero and finite temperatures. Besides a localization at certain impurity concentrations, the conductivity shows a maximum at small temperatures interpreted as onset of superconducting behaviour triggered by impurities. The Tan contact reveals the same universal behaviour as known from electron-electron scattering. 
\end{abstract}
\maketitle

\section{Introduction}

One-dimensional interacting Fermions have been investigated intensively due to their unusual properties. Examples of interacting one-dimensional systems range from large-scale structures like crystalline ion beams observed in high-energy storage rings \cite{SSH01,SSBH02} up to one-dimensional quantum wires \cite{PMC14} experimentally found in carbon nanotubes \cite{Saito98,Bockrath99,Ishii03,Shiraishi03}. Also edge states in quantum hall liquid \cite{Milliken96,Mandal01,Chang03}, semiconducting nanowires \cite{Schafer08,Huang01}, cold atomic gases \cite{Monien98,Recati03,Moritz05} and conducting molecules \cite{Nitzan03} are one-dimensional quantum systems.

Due to these versatile applications it is important to understand the ground state and transport properties of quantum wires. Frequently bosonization techniques in \cite{Lu77,Solyom79} and out of \cite{GGM10} equilibrium are employed which are based on the similar behaviour of long-distance correlations of Fermi and Bose systems \cite{H81a}. Exact solutions are available for Luttinger \cite{L63,LP74,ES07}, Tomonaga \cite{DL74}, and Gaudin-Yang models \cite{Ram17} of contact interaction by the Bethe ansatz \cite{EFGKK10,GBML13}. The ground-state properties of one-dimensional quantum wires have been analytically and numerically investigated \cite{LD11,LG16}. There the high-density expansion \cite{Loos13} of the ground-state energy was discussed in dependence on the width \cite{GADMP22} of the wire. The weak coupling corresponds to the high-density regime and the strong coupling regime to low densities due to the peculiar density dependence of the kinetic and interaction energy in one dimension \cite{GPS08}.

Green functions allow to describe interacting models beyond exactly solvable cases and in various approximations \cite{T67,VMKA08}. Especially the transition between Tomonaga-Luttinger and Fermi liquids can be investigated\cite{Sch77,Yo01}. For an overview about theoretical models see \cite{V94,G04,GV08}. 
With the help of the Green functions transport properties of correlated systems have been successfully described by the extended quasiparticle picture which relies on small scattering rates.
The limit of small scattering rates was first 
introduced by \cite{C66a} for highly degenerated Fermi liquids, 
later used in \cite{SZ79,KKL84} for equilibrium nonideal plasmas. The 
same approximation, but under the name of the generalised
Beth-Uhlenbeck approach, has been used by \cite{SR87,MR94} in
nuclear matter studies of the correlated density or in the kinetic 
equation for nonideal gases, \cite{BKKS96}. The resulting 
quantum kinetic equation unifies the achievements of transport in dense gases with the quantum transport of dense Fermi systems \cite{SLM96,MLS00,M17b}. 
In this kinetic equation the quasiparticle drift of Landau's
equation is connected with a dissipation governed by a nonlocal and non-instant
scattering integral in the spirit of Enskog corrections \cite{SLM98}. These corrections
are expressed in terms of shifts in space and time which characterize the
non-locality of the scattering process \cite{MLSK98}. In this way quantum
transport is possible to recast into a quasi-classical picture. The nonequilibrium quantum hydrodynamics resulting from this nonlocal kinetic theory can be found in \cite{M17,M17b}.

These successful perturbation approaches fail for one-dimensional electron systems since the Fermi surface disappears at T = 0 indicating the
breakdown of single-particle excitation turning them into collective ones \cite{SDC19}.
Quasiparticles are not developed at the Fermi energy \cite{DBG10} as seen e.g. in Hubbard Hamiltonian near half filling \cite{GVM93}. Nevertheless such excitations can show up eventually at the Luttinger surface where the Green functions have a zero at zero energy \cite{FA22}. 
Due to the absence of quasiparticles, expansions are necessary beyond the quasiparticle pole approximation. Unfortunately the expansion of small scattering rates relies on the finite behaviour of the reduced density at the Fermi energy and fails for non-Fermi liquids due to the divergence at the Fermi energy. Here we will suggest a twofold expansion as possible solution and a Pad\'e interpolation between both expansions. This will reproduce the spectral function and the reduced density or momentum distribution at the Fermi energy. The absence of Fermi surface is seen also in the momentum distribution (reduced density) which shows in one-dimensional Hubbard models \cite{Schu90,BVB90}, Gaudin-Yang model \cite{Ram17}, electron-impurity, and electron-phonon systems \cite{BS19} or Tomonaga models \cite{GS67} a continuous behaviour at the Fermi momentum controlled by the Luttinger behaviour. This reduced density is accompanied by a universal momentum asymptotic of $1/p^4$ reported of both free and harmonically trapped atoms for all values of the interaction strength \cite{OD03}. We will show that this continuous behaviour can be reproduced if the extended quasiparticle approximation is accompanied by the regulator of the suggested Pad\'e approximation.

For transport properties as integrals over the reduced density we will see that contribution of the Pad\'e approximations related to the interpolating function are negligible and the results of the extended quasiparticle picture remains. In this way the validity of the extended quasiparticle picture is extended beyond the originally derived perturbation of small scattering rates.

As illustration of this suggested Pad\'e regularization, we consider only scattering of electrons on impurities and assume that any electron-electron interaction is screened off. Similar models have been investigated in second Born approximation in \cite{BS19} with a screened Debye potential. Here we will even consider the simpler case of contact interaction but will discuss the Born and the T-matrix approximations \cite{M02}. The aim is not to describe the system as exact as possible but to illustrate the expansion scheme of the spectral function and how a two-term Pad\'e interpolation can reproduce the spectral function and reduced density matrix. This will repair the deficiency of the extended quasiparticle approximation.

The exact Bethe ansatz solution shows that an impurity is dressed up by surrounding particles and in the strongly attractive limit, it forms a dimer with inner Fermions in the Fermi
sea \cite{SZ19} similar to the highly imbalanced 1D Fulde-Ferrell-Larkin-Ovchinnikov (FFLO) state \cite{Fulde64,Larkin65}. In this respect it has been reported that quasi-one-dimensional superconductivity can be enhanced by disorder with screened Coulomb interaction \cite{PAC16}. Various other model systems have been found for quasi-one-dimensional superconductors \cite{Be11,WSLS12,He15}.  Some synthesis and fluctuation problems are overcome by building superconductors
from inhomogeneous composites of nanofilaments \cite{PHC16}. We will show the onset of superconductivity in a one-dimensional electron system interacting with impurities at finite momentum. It will be seen further-on that any approximation beyond Born, like e.g. the T-matrix summation, requires the inclusion of the correlated density due to scattering delay. The latter one leads to a correction of transport properties, e.g. the conductivity, due to the two-particle correlations. 

The outline is as follows. In the next chapter we briefly give the many-body scheme with the focus on the extended quasiparticle picture and discuss its failure due to the behaviour at the Fermi energy. This leads to the suggestion of a twofold expansion interpolated by a Pad\'e formula. In chapter III this twofold expansion is derived and the two first energy-weighted sum rules are proven to be completed independent on the interpolating function. In chapter IV the Pad\'e expression is applied successively to higher-order approximations starting first with a model of constant damping proceeding to a model of impurity scattering with contact  interaction in Born and T-matrix approximation. The influence of the bound-state poles are illustrated in the latter approximation. Two necessary amendments are found, the correction of energy argument of the selfenergy in order to reproduce the Fermi energy and the correlated density besides the free density. The reason of the latter is the delay time appearing in approximations with a dynamical vertex as two-particle correlation which is explained in chapter V. The Tan contact is given and compared to other approaches showing a universal behaviour. In chapter VI we discuss the effect of the delay time to the conductivity and give the finite temperature expression. It shows a localization at certain impurity densities and an enhancement at small temperatures as a possible onset of superconducting behaviour. Chapter VII summarizes and the appendix gives the Euler expansion of the $arctan$ function and the proof of cancellation of interpolation-function related terms for transport. This concludes the reasoning why the extended quasi-particle picture works also for correlated one-dimensional systems.  

\section{Many-body scheme}
\subsection{Extended quasiparticle approximation}

The Dyson equation for the causal propagator $G(1,2)=-i\langle {\cal T} a_1^+a_2 \rangle$ with time ordering ${\cal T}$ reads
\be
G(1,2)=G_0(1,2)+G_0(1,3) \Sigma(3,4)G(4,2)
\label{Dyson}
\ee
in terms of the Hartree-Fock propagator
\be
\left (i\frac{\partial}{\partial
t_1}\!+\!\frac{\nabla_1^2}{2m}\right ) G_0 (1,2)
\!-\!\Sigma^{HF}(1,3) G_0 (3,2)  =
\delta(1-2)
\label{g0}
\ee
with the time-diagonal Hartree-Fock self energy 
\be
\Sigma^{HF}(1,2)&=&\left [V(x_1\!-\!x_2) G(2,2^+)\!-\!V(x_1\!-\!x_2) G(1,2)\right ]\nonumber\\&&\times\delta(t_1\!-\!t_2),
\ee
and the correlation selfenergy $\Sigma(1,2)$. It is integrated about double occurring indices, $1=(\V x_1,t_1)$. The interaction potential is $V$ and $2^+$ signs an infinitesimal later time than $t_2$. 

Since we concentrate on equilibrium, all quantities are only dependent on the difference of coordinates. One Fourier transforms these difference coordinates into frequency/momentum. We will denote these transformed quantities by small letters. Nonequilibrium expressions can be found in \cite{M17b}.

The real part of the selfenergy is the Hilbert transform 
\be 
\sigma(\omega,q)={\rm Re}\, \sigma^R(\omega,q)=\int {d\bar \omega \over 2 \pi}{\gamma(q,\bar \omega)\over \omega-\bar \omega}
\label{Hilbert}
\ee
of the selfenergy spectral function
\be
\gamma=\sigma^> +\sigma^<=i(\sigma^R-\sigma^A)=-2 {\rm Im}\,  \sigma^R.
\label{Ga}
\ee
Both specifying the retarded selfenergy
\be
\sigma^R(\omega,k)=\sigma(\omega,k)-\frac i 2 \gamma(\omega,k)=\int {d\bar \omega \over 2 \pi}{\gamma(\bar \omega,k)\over \omega-\bar \omega+i\eta}.
\label{Sr}
\ee

The retarded propagator we obtain from (\ref{Dyson}) by Langreth/Wilkins \cite{LW72} rules leading to the same form of equation which are easily solved in equilibrium as
\be
g^R(\omega,k)={1\over \omega-\epsilon_k^0-\sigma^R(\omega,k)} 
\label{gr}
\ee
where we abbreviate in the following
\be
\varepsilon^0_k=\epsilon_k+\sigma^{\rm HF}_k.
\label{eps0}
\ee
The spectral function follows
\be
a(\omega,k)&=&g^>\pm g^<=i (g^R-g^A)
\nonumber\\
&=&{\gamma(\omega,k)\over  [\omega-\epsilon_k^0-\sigma(\omega,k)]^2+{\gamma(\omega,k)^2\over 4}}
\label{a}
\ee
for Bose/Fermi particles. 
The poles of (\ref{a}) at 
\be
\varepsilon_k=\varepsilon^0_k+\sigma(\varepsilon_k,k)
\label{pole}
\ee
describe the quasiparticle excitations by the selfconsistent energy $\varepsilon_k$  and $\gamma$ represents the quasiparticle damping. For question concerning the convergence of different many body expansions see \cite{Fa21}.

Within the extended quasiparticle picture we expand the spectral and correlation functions with respect to the order of damping
\ba
a^{\rm EQP}\!(\omega,k)={2\pi\delta(\omega\!-\!\varepsilon_k\!)
\over 1\!-\!{\partial \sigma(\omega,k)\over \partial \omega}}n_{\varepsilon_k}\!+\!
\gamma(\omega,k){\wp^\prime\over\omega\!-\!\varepsilon_k}\!+\!o(\gamma^2)
\label{as}
\end{align}
and correspondingly
\ba
g^<(\omega,k)={2\pi\delta(\omega\!-\!\varepsilon_k)
\over 1\!-\!{\partial \sigma(\omega,k)\over \partial \omega}}n_{\varepsilon_k}\!+\!
\sigma^<(\omega,k){\wp^\prime\over\omega\!-\!\varepsilon_k}\!+\!o(\gamma^2)
\label{gs}
\end{align}
with the derived principal value ${\wp^\prime\over\omega}=-\partial_\omega {\wp\over\omega}$ and the equilibrium distribution function $n_\omega$, i.e. the Fermi- or Bosefunction respectively.
The form (\ref{gs}) was presented with respect to small damping (scattering rate) expansion and with quasiparticle energies under the name of extended quasiparticle approximation in nonequilibrium \cite{SL94,SL95} and used for transport in impurity systems \cite{SLMa96,SLMb96}. The nonlocal kinetic theory finally is based on this expansion \cite{SLM96,MLS00,M17b}.

If we integrate (\ref{gs}) over the energy $\omega$ we get the connection between the reduced density matrix $\rho$ (momentum distribution) and the free (quasiparticle) distribution $n_{\varepsilon_k}$ as
\be
\rho(k)&=&n_{\varepsilon_k}+\int {d\omega \over 2 \pi} {\wp^\prime\over\omega\!-\!\varepsilon_k}\left [\sigma^<(\omega,k)-\gamma(\omega,k)n_{\varepsilon_k}\right ]
\nonumber\\
&=&n_{\varepsilon_k}+\int {d\omega \over 2 \pi} {\wp^\prime\over\omega\!-\!\varepsilon_k}\gamma(\omega,k)\left [n_\omega-n_{\varepsilon_k}\right ].
\label{red}
\ee
The quasiparticle distribution $n_{\varepsilon_k}$ is to be taken at the pole $\omega=\varepsilon_k$ of the spectral function (\ref{pole}) and the second line of (\ref{red}) is valid in equilibrium  where we have $\sigma^<(\omega,k)=n_\omega \gamma(\omega,k)$.

For Fermions in the ground state, $n_\epsilon=\Theta(\epsilon_F-\epsilon)$, observing that 
\be
n_\omega\!-\!n_\varepsilon=\Theta(\varepsilon \!-\! \epsilon_F)\Theta(\epsilon_F-\omega)\!-\!\Theta(\epsilon_F\!-\!\varepsilon)\Theta(\omega\!-\!\epsilon_F)
\label{relat}
\ee
we can further rewrite (\ref{red}) into
\ba
\rho(k)=&\Theta(\epsilon_F-\varepsilon_k)
+\Theta(\varepsilon_k \!-\! \epsilon_F)\int\limits_{-\infty}^{\epsilon_F} {d\omega \over 2 \pi} {\wp^\prime\over\omega\!-\!\varepsilon_k}\gamma(\omega,k)
\nonumber\\
&-\Theta(\epsilon_F\!-\!\varepsilon_k)\int\limits^{\infty}_{\epsilon_F} {d\omega \over 2 \pi} {\wp^\prime\over\omega\!-\!\varepsilon_k}\gamma(\omega,k).
\label{red1}
\end{align}
This shows that if we approach the Fermi energy from above $\varepsilon_k=\epsilon_F+0$ we have the value of the second term in (\ref{red1})
\be 
z_+=\int\limits_{-\infty}^{0} {d\omega \over 2 \pi} {\wp^\prime\over\omega}\gamma(\omega+\epsilon_F,k_F)
\label{zp}
\ee
and approaching from below $\varepsilon_k=\epsilon_F-0$ the step function is reduced by a factor given by the third term in (\ref{red1})
\be 
z_-=1-\int\limits^{\infty}_{0} {d\omega \over 2 \pi} {\wp^\prime\over\omega}\gamma(\omega+\epsilon_F,k_F).
\label{zm}
\ee 
Together one sees a jump at the Fermi energy of
\be
z_--z_+&=&1-\int\limits^{\infty}_{-\infty} {d\omega \over 2 \pi} {\wp^\prime\over\omega}\gamma(\omega+\epsilon_F,k_F)
\nonumber\\
&=&1+\left . \partial_\omega \sigma(\omega,k_F)\right |_{\omega=\epsilon_F}
\ee
i.e. just the wave-function renormalization as the factor of the pole in (\ref{as}). This typical Fermi-liquid behaviour is present as long as the imaginary part of selfenergy, the damping $\gamma$ of the quasiparticles, vanishes at the Fermi surface
\be
\lim\limits_{\omega\to 0}\gamma(\omega+\epsilon_F,k_F)=0
\ee
in order to render (\ref{zp}) and (\ref{zm}) finite as necessary condition. If this requirement is not fulfilled then the integrals diverge and the perturbation theory breaks down. This we explore in the following section.

\subsection{Problems with the extended quasiparticle approximation for non-Fermi liquids}

\subsubsection{Model of constant quasiparticle damping}
\begin{figure}[h]
\includegraphics[width=8.5cm]{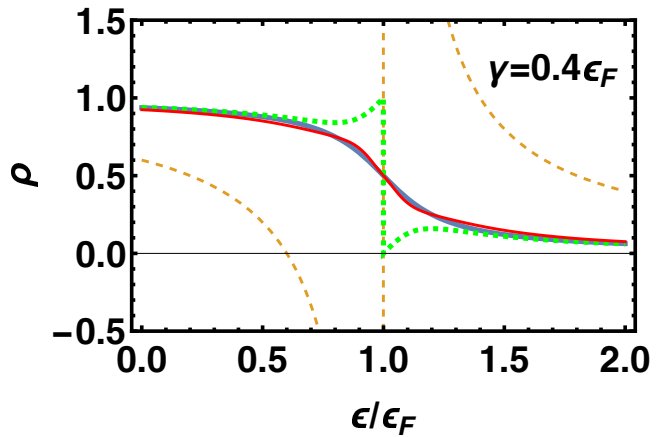}
\caption{The reduced density matrix (\ref{arct}) (thick) together with the extended quasiparticle picture (dashed) and the Euler expansion (\ref{Euler}) (dotted) and the Pad\'e interpolation (\ref{inter}) (thin red line) visibly not distinguishable from the exact reduced density matrix. 
\label{aspec}}
\end{figure}

The problem of the perturbation expansion and resulting extended quasiparticle picture is best understood by a model of constant quasiparticle damping $\gamma(\omega,k)=\gamma$ and consequently $\sigma=0$ and $\varepsilon_k=\epsilon_k$ due to (\ref{Hilbert}).
The exact reduced density matrix reads
\be
\rho(k)&=&\int\limits_{-\infty}^\infty {d\omega\over 2 \pi} a(\omega,k) n_\omega=\int\limits_{-\infty}^{\epsilon_F} {d\omega\over 2 \pi} {\gamma\over (\omega-\epsilon_k)^2+{\gamma^2\over 4}} 
\nonumber\\
&=&\frac 1 2 +\frac 1 \pi {\rm arctan}{\epsilon_F-\epsilon_k\over \gamma/2}.
\label{arct}
\ee
The extended quasiparticle approximation as the expansion with respect to $\gamma$ of (\ref{red1}) would be
\be
\rho(k)=\Theta(\epsilon_F-\varepsilon_k)-\gamma {\wp \over \epsilon_F-\varepsilon_k}
\ee
and shows a divergence at the Fermi energy. This puzzling failure can be seen from the exact expression (\ref{arct}). The $arctan$ has a cut such that the expansion near the Fermi energy, which means at small values of $\epsilon_F-\epsilon$, is different from the expansion of small damping $\gamma$,
\be 
{\rm arctan}(x)=\left \{\begin{array}{l}
x+o(x^2)
\cr
\frac \pi 2 {\rm sig} (x) -{1\over x}+o({1\over x^2})
\end{array}\right . 
\ee
which explains why the expansion diverges at $x=2(\epsilon_F-\varepsilon_k)/\gamma\to 0$, seen in figure~\ref{aspec}.

A somewhat better expansion which does not diverge at the Fermi energy was found by Euler in 1755, see appendix~\ref{Eulerderv},
\be 
{\rm arctan} (x)=\left \{\begin{array}{l}
{x\over 1+x^2}+o(x^2)
\cr
\frac \pi 2 {\rm sig} (x) -{x\over 1+ x^2}+o({1\over x^2})
\end{array}\right . 
\label{Euler}
\ee
which does not diverge at $x=0$ but remains discontinuous as seen in figure~\ref{aspec}.
This problem of the cut cannot be circumvented. A 
solution could be to use the Euler-expansion (\ref{Euler}) and to interpolate between both expansions
\be 
{\rm arctan} (x\!)\!=\!
{x\over 1\!+\!x^2} f(x)\!+\!
\left [\frac \pi 2  {\rm sig} (x)\!-\!{x\over 1\!+\! x^2}\right ][1\!-\!f(x)]
\label{inter}
\ee
with a Pad\'e function $f(x)$ approaching $1$ for small $x$ and $0$ for large $x$. We choose here $f(x)=1/(1+x^2)^2$ which gives an excellent agreement with the exact result as seen in figure~\ref{aspec}. Using $\Theta(x)=[1+{\rm sig} (x)]/2$, the Pad\'e interpolation (\ref{inter}) translates into the reduced density matrix for (\ref{arct})
\ba
\rho^{\rm Pade}(k)\!=&\left (\!\frac 1 2 \!+\!\frac 1 \pi{x\over 1\!+\!x^2}\!\right )\! f(x)
\nonumber\\&\!+\!
\left (\!\Theta(x) \!-\!\frac 1 \pi {x\over 1\!+\! x^2}\right )[1\!-\!f(x)]
\label{interrho}
\end{align}
with $x=2 (\epsilon_F-\epsilon_k)/\gamma$.

\section{Pad\'e approximation for the spectral function}
\subsection{Twofold expansion of spectral function}

The obviously working idea of Pad\'e approximation of the last chapter we derive now by a systematic expansion of the spectral function. For this purpose we observe that the Euler formula (\ref{Euler}) is obtained if we replace formally the standard Taylor expansion by a finite value form
\be
g(x)\approx g(0)+g'(0) x \longrightarrow g(x)\approx g(0)+g'(x) x
\label{expand}
\ee
with $g(x)={\rm arctan}(x)$.

In the spectral function (\ref{a}) we have now two expansions, one with respect to small $\gamma$ and one with respect to deviations from the Fermi level 
$\varepsilon_k-\epsilon_F$
with $\varepsilon_k$ the quasiparticle energy as solution of (\ref{pole}). Both expansions have to be interpolated according to (\ref{inter}) due to the cut. We translate now this expansion for (\ref{arct}) into an expansion for the spectral function in such a way to reproduce (\ref{inter}). Applying the rule (\ref{expand}) for both expansions, $\gamma$ and $\varepsilon_k-\epsilon_F$, we suggest
\ba
a(\omega,k)\approx &a^{\rm Pade}(\omega,k)=a_{\rm f}(\omega,k)\, f\!\!\left (x\right )\!+\!a_{1\!-\!{\rm f}}(\omega,k)\,\left [1\!-\!f\!\left (x\right )\right ]
\label{pade}
\end{align}
with $x=2 (\epsilon_F\!-\!\varepsilon_k)/\gamma(\epsilon_F,k)$. The small scattering-rate expansion of the standard extended quasiparticle approximation (\ref{as}) is denoted as
\ba
a_{1\!-\!{\rm f}}(\omega,k)=&2\pi \delta[\omega\!-\!\epsilon_k\!-\!\sigma(\omega)]\nonumber\\
&-\gamma(\omega,k)\partial_{\omega} {\omega \!-\!\varepsilon_k\over (\omega\!-\!\varepsilon_k)^2\!+\!{\gamma^2(\varepsilon_k,k)\over 4}}.
\label{EQPA1}
\end{align}
Here we have extended the last term by the damping according to (\ref{expand}). The additional part in (\ref{pade})
\ba
a_{\rm f}(\omega,k)=&
{\gamma(\epsilon_F,k)\over (\omega-\varepsilon_F)^2\!+\!{\gamma^2(\epsilon_F,k)\over 4}}\nonumber\\
&-(\epsilon_k\!-\!\epsilon_F)\partial_{\omega} 
{\gamma(\epsilon_F,k)\over (\omega-\varepsilon_F)^2\!+\!{\gamma^2(\epsilon_F,k)\over 4}}
\label{EQPA2}
\end{align}
describes now the expansion with respect to the deviation from the Fermi energy.  The subtle difference has to be noted in the pre-factor of the second term being the deviation of the Hartree-Fock energy $\epsilon_k-\epsilon_F$ from the Fermi energy and not the quasiparticle energy $\varepsilon_k-\epsilon_F$. This is not visible from the constant model so far where both expressions are identical. That this extension is correct we will convince ourselves now by various sum rules and the comparison with the reduced density (\ref{interrho}) for successively more complicated models in chapter IV.  

\subsection{Reduced density matrix}
First it is instructive to see how the reduced density matrix (\ref{red}) appears from the Pad\'e approximation (\ref{pade}). One has as exact spectral expression
\ba
\rho(k)\!=\!
\int\limits_{-\infty}^{\infty}\!\!{d \omega\over 2\pi} a(\omega,k)n_\omega=\!\int\limits_{-\infty}^{\epsilon_F}\!\!{d \omega\over 2\pi} a(\omega,k).
\label{red2}
\end{align}
The integral over the $\delta$-function as first part in (\ref{EQPA1}) reads
\ba
&\int\limits_{-\infty}^0\!\! d\bar \omega \delta [\bar \omega\!+\!\epsilon_F\!-\!\epsilon_k\!-\!\sigma(\bar \omega\!+\!\epsilon_F)]=\!
\int\limits_{-\infty}^0\!\! d\bar \omega
           {\delta (\bar \omega\!-\!\bar \varepsilon_k)\over 1\!-\!\partial_{\bar \omega} \sigma}
           \nonumber\\
&\approx\!
           \Theta(\epsilon_F-\varepsilon_k)
           \left [1\!+\!\int{d\omega\over 2 \pi} \gamma(\omega,k)\partial_\omega
             {\omega\!-\!\varepsilon_k\over (\omega\!-\!\varepsilon_k)^2+{\gamma^2(\varepsilon_k,k)\over 4}}\right ]
\label{delt}
\end{align}
where we used (\ref{Hilbert}). The approximation includes the expansion of the denominator in accordance with first-order damping and the subsequent extension of the principal value by the damping according to (\ref{expand}). This combines now with the second part of (\ref{EQPA1}) to yield
\ba
&\rho_{1\!-\!{\rm f}}(k)=\int\limits_{-\infty}^{\epsilon_F}\!\!{d \omega\over 2\pi} a_{1\!-\!{\rm f}}(\omega,k)\! =\!
\Theta(\epsilon_F-\varepsilon_k)
\nonumber\\&
+\int \!\!{d\omega \over 2 \pi} \gamma(\omega,k)
\left [\Theta(\epsilon_F\!-\!\varepsilon_k)\!-\!\Theta(\epsilon_F\!-\!\omega)\right ]\partial_\omega \tilde b(\omega,k)
\label{t1}
\end{align}
where the last line can be rewritten with the help of (\ref{relat})
\ba
\!\left (\!\!
\Theta(\epsilon_F\!-\!\varepsilon_k)\!\!\int\limits_{\epsilon_F}^\infty \!\!{d\omega \over 2 \pi}
\!-\!
\Theta(\varepsilon_k\!-\!\epsilon_F)\!\int\limits^{\epsilon_F}_{-\infty} \!\!{d\omega \over 2 \pi}
\!\right )\!
\gamma(\omega,k)
\partial_\omega \tilde b(\omega,k)
\end{align}
with the abbreviation
\be
\tilde b(\omega,k)={\omega\!-\!\varepsilon_k\over (\omega\!-\!\varepsilon_k)^2\!+\!{\gamma^2(\varepsilon_k,k)\over 4}}.
\label{tb}
\ee
For the additional part (\ref{EQPA2}) we can perform the frequency integral to get
\ba
&\rho_{\rm f}(k)=\int\limits_{-\infty}^{\epsilon_F}\!\!{d \omega\over 2\pi} a_{f}(\omega,k)
\nonumber\\
&=\left [ \frac{1}{ \pi} {\rm arctan}{\omega-\epsilon_F\over \gamma(\epsilon_F)} -{(\epsilon_k-\epsilon_F)\over 2 \pi} {\gamma(\varepsilon_k)\over 
 (\omega-\epsilon_F)^2+{\gamma(\epsilon_F)^2\over 4}}\right ]_{-\infty}^{\omega=\epsilon_F}
\label{t2}
\end{align}
where the above and below integration limits have to be inserted depending on the damping which might be zero.

As an intermediate check, adding both terms (\ref{t1}) and (\ref{t2}) yields exactly (\ref{interrho}) with the Pad\'e approximation (\ref{inter}) observing that here in this model we have $\gamma(\omega,k)=const$ and $\sigma=0$ and consequently $\varepsilon_k=\epsilon_k$.

The form of Pad\'e approximation suggested in (\ref{pade}) has a general validity for more refined models, e.g. also for electron-electron scattering since it relates to any selfenergy. As illustration we restrict to impurity models in the next chapters. To convince ourselves about the larger validity we first proof that the sum rules are completed independent of the used models.

\subsection{Sum rules}
We check the frequency sum rule valid for any model of the selfenergy. We can perform the trivial frequency integrals directly analogously to (\ref{t1}) and (\ref{t2}) but with the upper integration limit up to infinity to obtain
\ba
\int{d\omega \over 2 \pi } a_{1\!-\!{\rm f}}(\omega,k)=\int{d\omega \over 2 \pi } a_{f}(\omega,k)=1
\label{inta}
\end{align}
which means
\ba
\int{d\omega \over 2 \pi } a^{\rm Pade}(\omega,k)=[1-f(x)]+f(x)=1
\label{sumr}
\end{align}
is completed independent of the interpolation function $f(x)$ where we abbreviated $x=2 (\epsilon_F\!-\!\varepsilon_k)/\gamma(\epsilon_F,k)$.

Even the first energy-weighted sum rule is also fulfilled which already the extended quasiparticle approximation does \cite{M17b}. To see this we consider the integrals over (\ref{EQPA1}) and (\ref{EQPA2}) separately. Analogously to (\ref{t1}) we get with abbreviation (\ref{tb})
\ba
&\int{d \omega\over 2 \pi} \omega a_{1\!-\!{\rm f}}(\omega,k)=\varepsilon_k \!+\!\int\!\!{d\omega \over 2 \pi }[\varepsilon_k\!-\!\omega]\,\gamma(\omega,k)\,\partial_{\omega}\tilde b(\omega,k)
\nonumber\\&
\approx
\varepsilon_k+\int\!\!{d \omega \over 2 \pi }(\varepsilon_k-\omega)\gamma(\omega,k)\partial_\omega\!
{1\over \omega\!-\!\varepsilon_k}+o(\gamma^3)
\nonumber\\&
=\varepsilon_k+\int\!\!{d\bar \omega \over 2 \pi }{\gamma(\omega,k)\over \bar \omega-\!\varepsilon_k}=\varepsilon_k-\sigma(\varepsilon_k).
\end{align}
The integral over (\ref{EQPA2}) reads
\ba
&\int{d \omega\over 2 \pi} \omega a_{\rm f}(\omega,k)
\nonumber\\
&=\epsilon_F \!-\!\int{d \omega\over 2 \pi} \omega (\epsilon_k\!-\!\epsilon_F)\partial_\omega {\gamma(\epsilon_F,k)\over (\omega\!-\!\epsilon_F)^2+{\gamma^2(\epsilon_F.k)\over 4}}
\nonumber\\&
=\epsilon_F \!+\!\int{d \omega\over 2 \pi} (\epsilon_k\!-\!\epsilon_F){\gamma(\epsilon_F,k)\over (\omega\!-\!\epsilon_F)^2+{\gamma^2(\epsilon_F,k)\over 4}}
=\epsilon_k.
\end{align}
Adding both parts we obtain the correct first energy-weighted sum rule
\ba
&\int{d \omega\over 2 \pi} \omega a^{\rm Pade}(\omega,k)=(\varepsilon_k-\sigma)(1-f(x))+\epsilon_k f(x)=\epsilon_k.
\label{sumw}
\end{align}
It is remarkable that both sum rules are fulfilled for any interpolating function $f[2(\epsilon_F- \varepsilon_k)/\gamma(\epsilon_F,k)]$ and any model for the selfenergy. This suggests already that for transport, i.e. integrated forms of the reduced density, the extended quasiparticle term $\rho_{\rm 1-f}$ might be sufficient which we will proof indeed in appendix~\ref{cancel} which is valid for any approximation of self energy. Therefore the Pad\'e approximation presented here is also valid for electron-electron scattering though only impurity scattering examples are chose in the following for illustration.

\section{Examples of Application}

Here we will compare the reduced density matrix from the integral over the spectral function (\ref{red2}) with the analytical Pad\'e approximations (\ref{pade})-(\ref{EQPA2}). This will be illustrated on two different many-body models, the Born and T-matrix approximation. For further models see e.g. \cite{Fa19}.

\subsection{Impurity scattering in Born approximation}
\subsubsection{Integrated spectral function}
We will now test the Pad\'e interpolation (\ref{pade})-(\ref{EQPA2})  with an exactly integrable model of frequency-dependent selfenergy. In Born approximation the scattering of particles with mass $m$ on impurities of density $n_i$ interacting by contact interaction $V_q=V_0$ the imaginary part of selfenergy or quasiparticle damping (\ref{Ga}) reads
\be
\gamma(\omega)&=&n_is \int{d q\over 2 \pi\hbar}V_q^2 2\pi \delta\left (\omega-{(k+q)^2\over 2m}\right )
\nonumber\\
&=&g{\Theta(\omega)\over \sqrt{\omega}}
\label{ig}
\ee
with the interaction constants  
\be
g&=&{s n_i V_0^2 4 m^2\over \hbar k_F^3}= {n_i\over n_F}{4 s^2\over \pi} b\nonumber\\
b&=&{\hbar^2\over k_F^2a_B^2}=r_s^2{s^4\over \pi^2}.
\label{gb}
\ee
In the following we understand all energies, $\omega, \gamma, \sigma$ etc, in units of Fermi energy and the momenta $k$ in units of Fermi momentum given by the free-particle density $n_F$  as $k_F=n_F \hbar \pi/s$ where we denote the spin-degeneracy by $s$. The interaction strength we express in terms of a Bohr-radius-equivalent $V_0=\hbar^2/ma_B$ which allows to discuss charged and neutral impurities on the same footing. The Bruckner parameter $r_s$ is the ratio of inter-particle distance $d=1/ns$ to this Bohr radius  $r_s=d/a_B$. 

In Born approximation we have only a dependence on the single parameter $g$. In T-matrix approximation presented in the next chapter we will see the dependence on both parameters $g$ and $b$ independently, i.e. the interaction strength and the impurity density.

\begin{figure}[h]
\includegraphics[width=8.5cm]{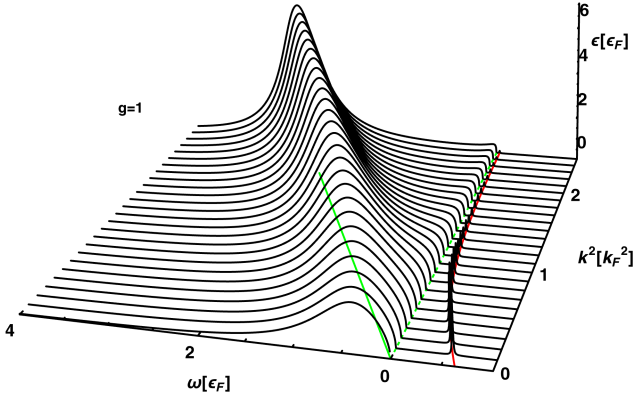}
\caption{The spectral function (\ref{a}) in Born approximation (\ref{Born}). The maxima at the quasiparticle energy (\ref{eps}) are indicated in the $\omega-k$ area as green line. The sharp bound states at negative energy are artificially broadened to make them visible. 
\label{awk}}
\end{figure}

The corresponding real part of self energy reads according to (\ref{Hilbert})
\be
\sigma(\omega)=-g{\Theta(-\omega)\over 2\sqrt{-\omega}}
\label{is}
\ee
as it was used in Eq. 30 of \cite{BS19} with Fermi liquid parameters.
The spectral function (\ref{a}) for this Born approximation can then be written with (\ref{ig}) and (\ref{is}) introducing in (\ref{gr})
\be
a(\omega)=\left \{
\begin{array}{ll}
  {{g\over \sqrt{\omega}}\over (\omega-k^2)^2+{g^2\over 4\omega}}&\omega>0
  \cr
  2\pi \delta(\omega-k^2+{g\over 2\sqrt{-\omega}})=2\pi {\delta(\omega+q_0^2)\over 2 q_0+{g\over 2 q_0^2}}&\omega<0
\end{array}
\right .
\label{Born}
\ee
where $q_0$ is the real solution of
\be
q_0^2+k^2-{g\over 2 q_0}=0
\label{q0}
\ee
and describes a sharp bound-state pole at negative energies due to vanishing damping.
The selfenergy (\ref{is}) results into the quasiparticle energy
\be
\varepsilon_k=\left \{\begin{array}{ll}k^2 & \omega>0\cr -q_0^2 &\omega<0\end{array}\right .
\label{eps}
\ee
and are the maxima of the spectral function as seen in figure~\ref{awk}. They follow the quasiparticle $\omega=k^2$ line but with a finite damping. In contrast, the negative bound-state poles at $\omega=-q_0^2$ are sharp. 

\begin{figure}[h]
  \includegraphics[width=4.2cm]{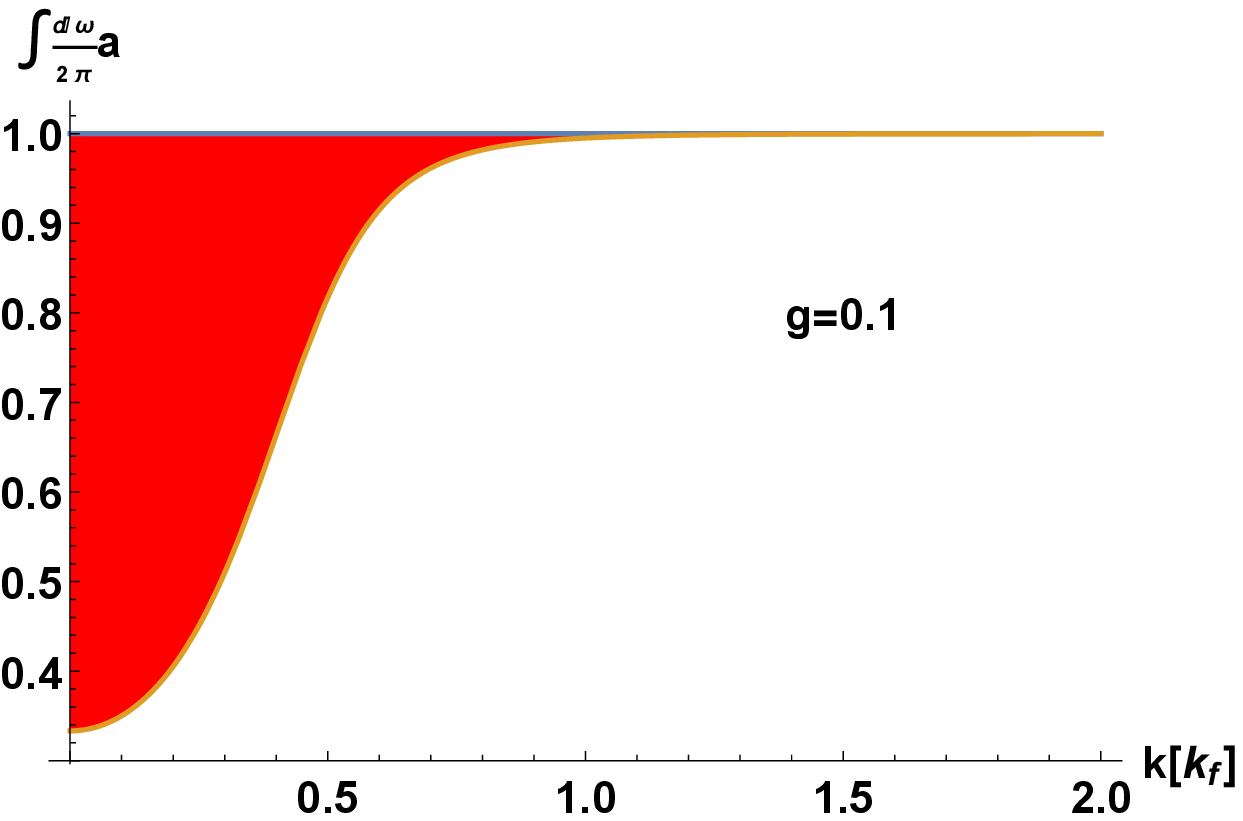}\includegraphics[width=4.2cm]{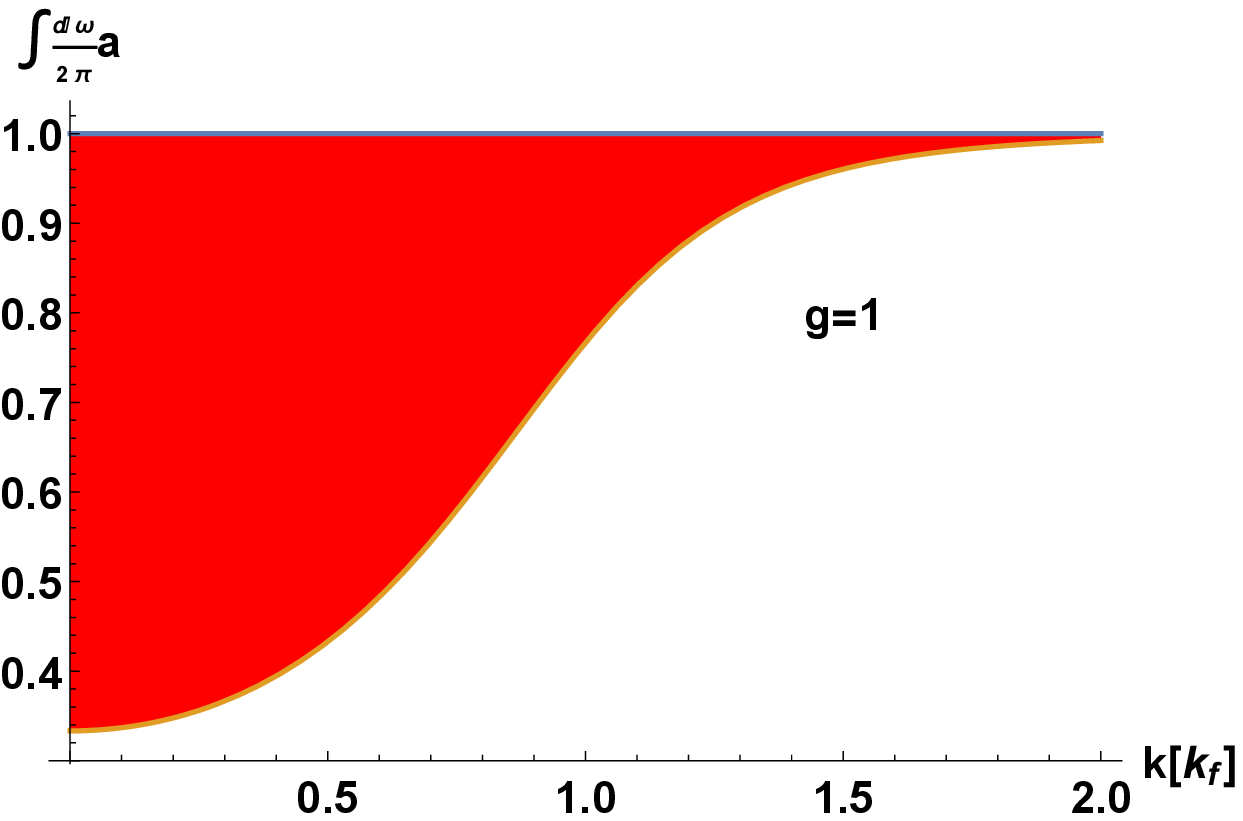}\\
  \includegraphics[width=4.2cm]{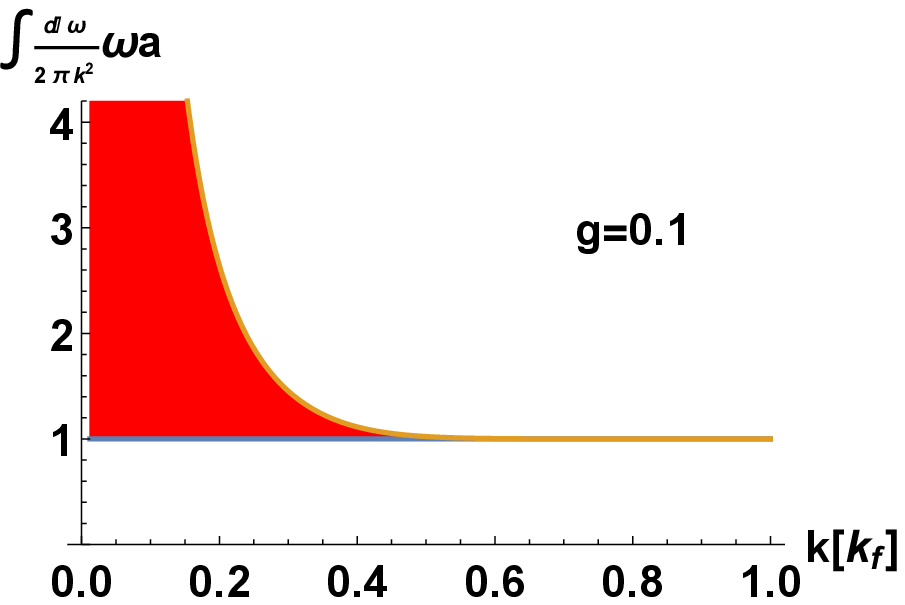}\includegraphics[width=4.2cm]{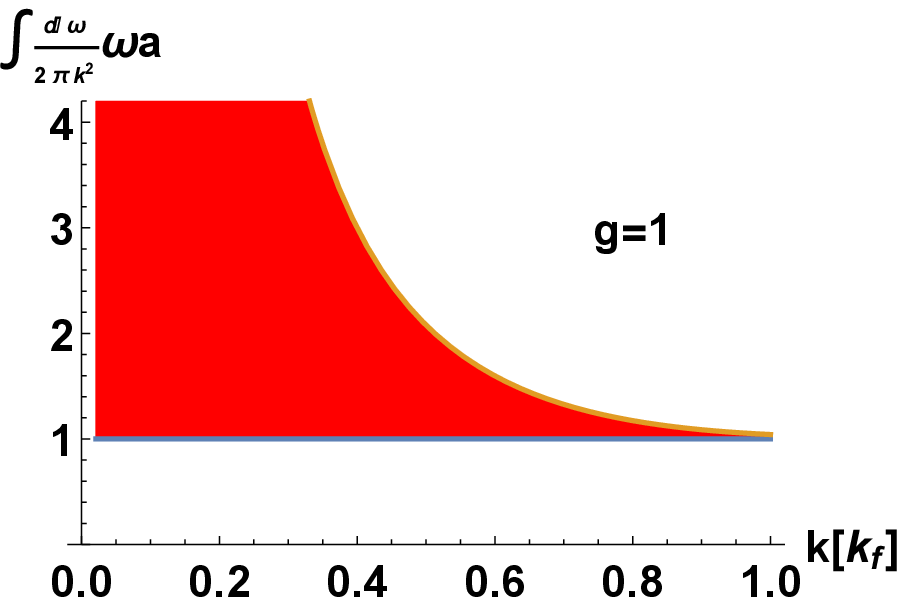}
\caption{The frequency sum rule (\ref{sumr}) (above) and the energy-weighted sum rule (\ref{sumw}) (below) for the impurity scattering model in Born approximation (\ref{Born}) for two different impurity couplings. The red area indicates the contribution of the bound-state pole (\ref{q0}) at negative frequencies. 
\label{aw}}
\end{figure}

The sum rules (\ref{sumr}) and (\ref{sumw}) are illustrated in figure~\ref{aw}. The substantial contribution of the bound-state pole to complete the sum rules  is visible.

\begin{figure}[h]
\includegraphics[width=4.2cm]{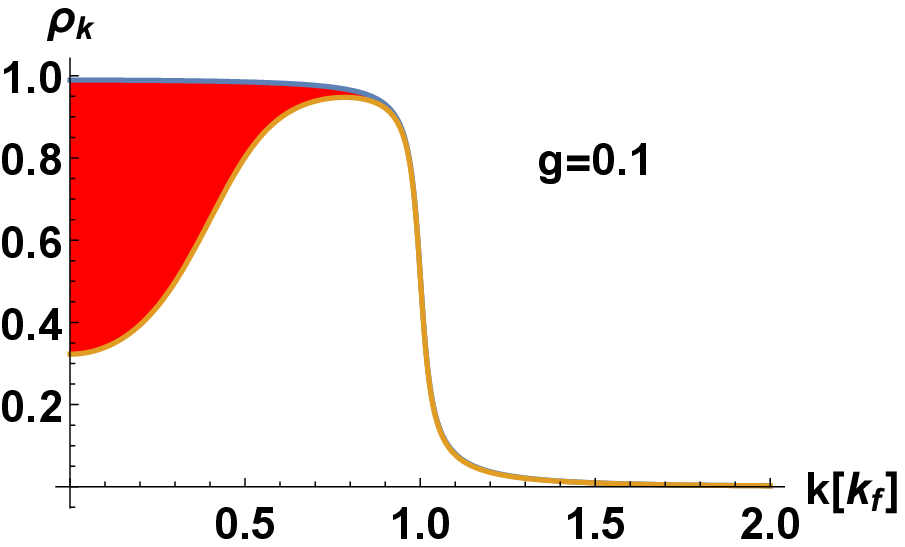}\includegraphics[width=4.2cm]{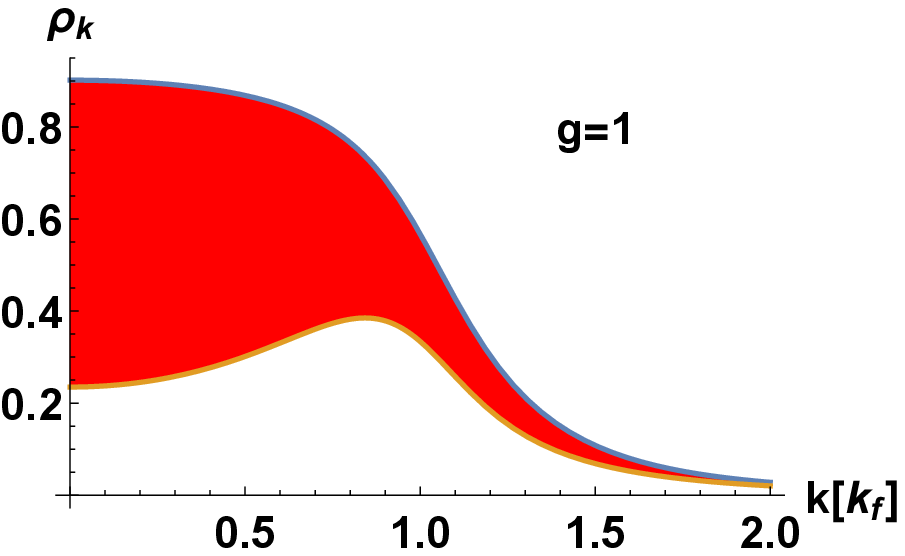}
\caption{The reduced density matrix from the integral (\ref{red2}) for the impurity scattering model in Born approximation (\ref{Born}) for two different impurity couplings. The red area indicates the contribution of the bound-state pole at negative frequencies. 
\label{rho}}
\end{figure}

The reduced density matrix (\ref{red2}) can be obtained analytically though the integral about positive frequencies is somewhat lengthy. In figure~\ref{rho} we plot the reduced density from spectral function (\ref{red2}) for two different coupling constants and illustrate the influence of the bound-state pole at negative frequencies. Its contribution is quite remarkable. 

\subsubsection{Pad\'e approximation}

The Pad\'e approximation (\ref{t1}) reads now for the impurity model (\ref{is}) and (\ref{ig})
\ba
&\rho_{1\!-\!{\rm f}}(k)=\Theta(1-k^2)
+\nonumber\\&-{g\over 2 \pi}
\!\left (\!\!
\Theta(1\!-\!k^2)\!\!\int\limits_1^\infty 
\!-
\Theta(k^2\!-\!1)\!\int\limits^1_{0} 
\right )\!\!{dq\over q}
\partial_q{q^2\!-\!k^2\over (q^2\!-\!k^2)^2\!+\!{g^2
    \over 4 q}}
\label{r1f}
\end{align}
with an elementary integral. The second part (\ref{t2}) becomes
\ba
\rho_{\rm f}(k)&={1\over 1+{g\over 4 q_0^3}}+\frac{1}{ \pi} {\rm arctan}{2\over g}
\nonumber\\&-{(k^2-1) g\over 2 \pi} \left [ {1\over 
 (1-k^2)^2+{g^2\over 4}}-{1\over 
 (k^2)^2+{g^2\over 4}}\right ]
\label{rf}
\end{align}
where we had to break the integration over frequency into parts larger and smaller zero whereby the latter one leads to the bound-state pole contribution.

\begin{figure}
  \includegraphics[width=8.5cm]{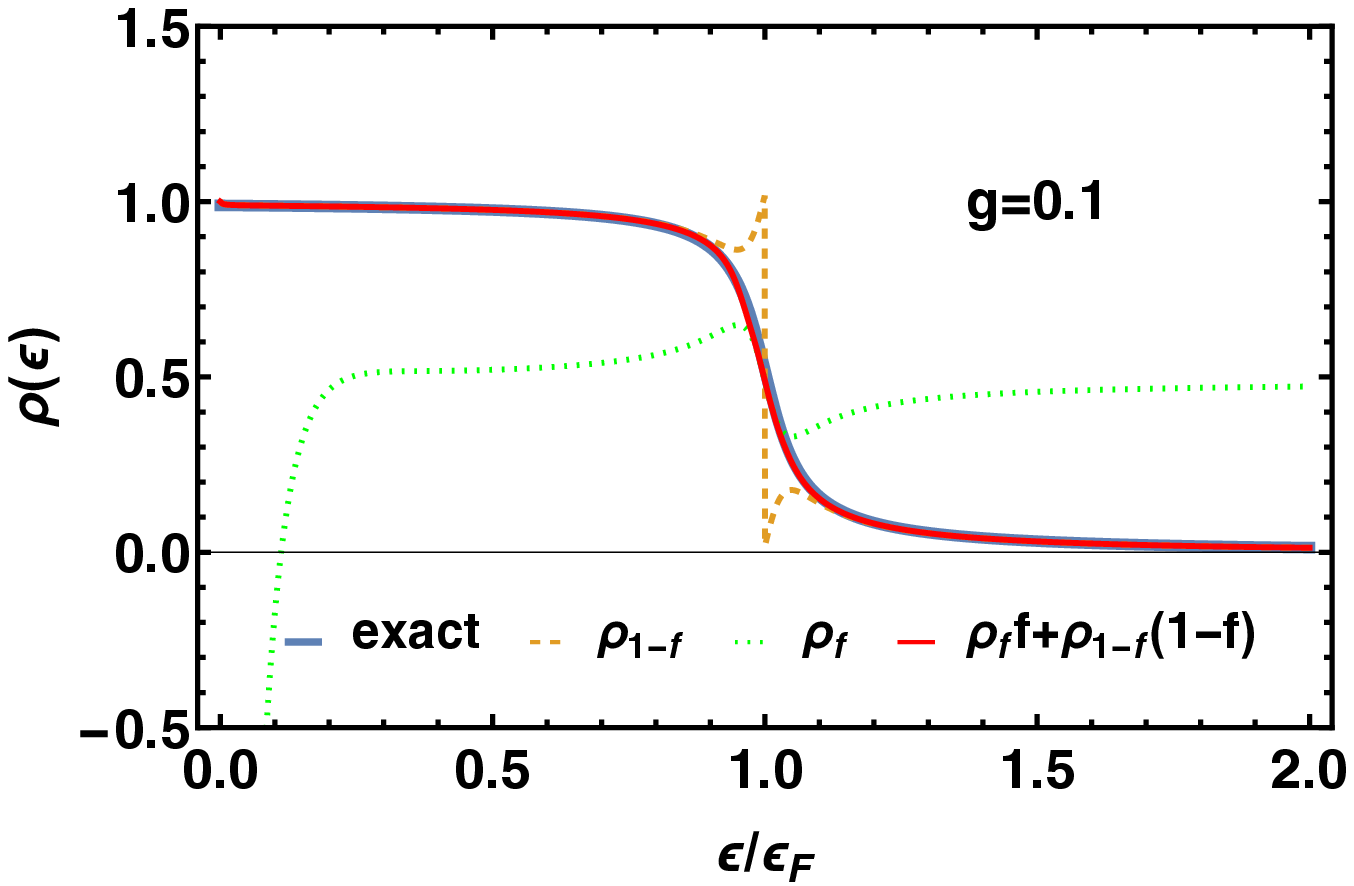}
  \includegraphics[width=8.5cm]{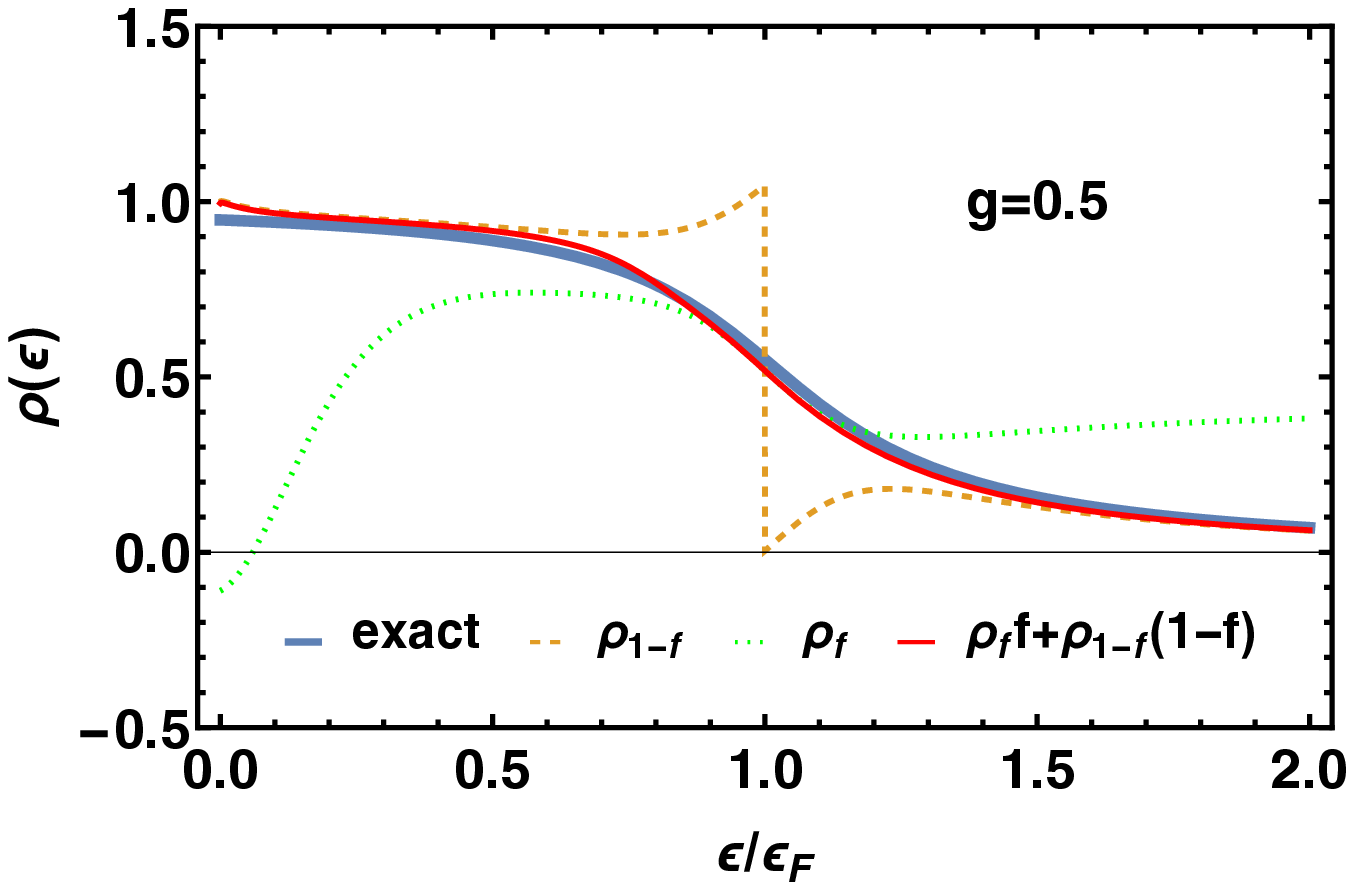}
\caption{The reduced density matrix by integration of the spectral function (\ref{red2}) in Born approximation (\ref{Born}) together with its Pad\'e approximation and the contributing parts (\ref{r1f}) and (\ref{rf}) for two different impurity couplings.  
\label{rhof}}
\end{figure}

In figure~\ref{rhof} we can see the reduced density matrix from (\ref{Born}) together with the Pad\'e approximation (\ref{pade}). Though the contributing parts (\ref{r1f}) and (\ref{rf}) are deviating strongly out of the area of their expansion range, the Pad\'e-weighted form shows a good approximation for different impurity couplings. Basically the role of the additional $\rho_F$ is to cure the jump of $\rho_{1-f}$ at the Fermi energy.

\subsection{Impurity scattering in Ladder approximation}
\subsubsection{Integrated spectral function}
Next we sum the ladder diagrams which means to solve the equation for the retarded $T$-matrix \cite{M02}
\be
T^R(\omega)&=&V+V\int{dk\over 2\pi \hbar} {T^R(\omega)\over \omega-{k^2\over 2 m}+i0}
\nonumber\\
&=&{\hbar^2\over m a_B}\left [ 1+\sqrt{b}\left ({\Theta(-\omega)\over \sqrt{-\omega}}+i {\Theta(\omega)\over \sqrt{\omega}} \right )\right ]^{-1}
\ee
with $b$ of (\ref{gb}). The correlation part of the selfenergy
simplifies from two-particle scattering towards impurity scattering used here
\be
\sigma^<(\omega,k)&=&s\sum\limits_{p,q}|T^R(\omega\!+\!\varepsilon_p^i,k,p)|^2 2\pi \delta(\omega\!+\!\varepsilon_p^i\!-\!\varepsilon_{k\!+\!q}\!-\!\varepsilon_{p\!-\!q}^i)\nonumber\\&&\times n_{k+q}n_{p-q}^i(1-n_p^i)
\nonumber\\
&=&n_is \sum\limits_{q}|T^R(\omega)|^2 2\pi \delta(\omega+\varepsilon_p^i-\varepsilon_{q})n_{q}
\nonumber\\
&=&{g\over \sqrt{\omega}+{b\over \sqrt{\omega}}}\Theta(\omega)\Theta(\epsilon_F-\omega)
\label{ssmall}
\ee
and $\sigma^>$ is obtained by interchanging $n\leftrightarrow 1-n$. The resulting imaginary (\ref{Ga}) and real (\ref{Hilbert}) parts of the selfenergy read
\be
\gamma(\omega)&=&\sigma^>+\Sigma^<=i(\sigma^R-\Sigma^A)= {g\over \sqrt{\omega}+{b\over \sqrt{\omega}}}\Theta(\omega)
\nonumber\\
\sigma(\omega)&=&-{g(\sqrt{b}-\Theta(-\omega)\sqrt{-\omega})\over 2(b+\omega)}
\label{tmat}
\ee
and compared to the Born approximation (\ref{ig}) and (\ref{is}) they deviate by the parameter $b$ given in (\ref{gb}). The expressions for Born and noncrossing approximations can be found in \cite{HuS93}.

\begin{figure}[h]
\includegraphics[width=10cm]{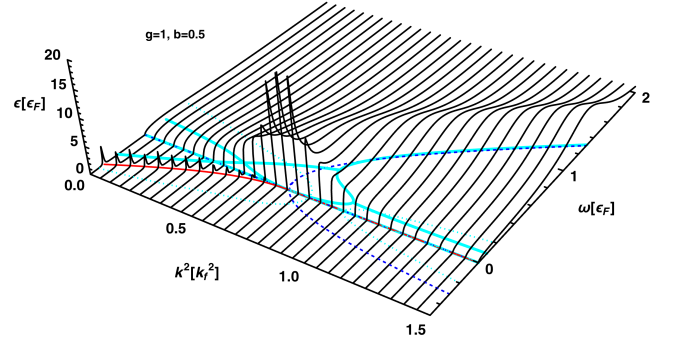}
\caption{The spectral function (\ref{a}) in T-matrix approximation (\ref{tmat}). The maxima/minima at the quasiparticle energy (\ref{eps}) are indicated as thick line in the $\omega-k$ area and their corresponding imaginary parts as dotted line. The zero of the quasiparticle dispersion (\ref{e12}) is plotted as dashed line. The sharp bound states at negative energy are artificially broadened. 
\label{ep3D}}
\end{figure}

The spectral function (\ref{a}) is plotted in figure~\ref{ep3D}. 

\begin{figure}[h]
\includegraphics[width=8.5cm]{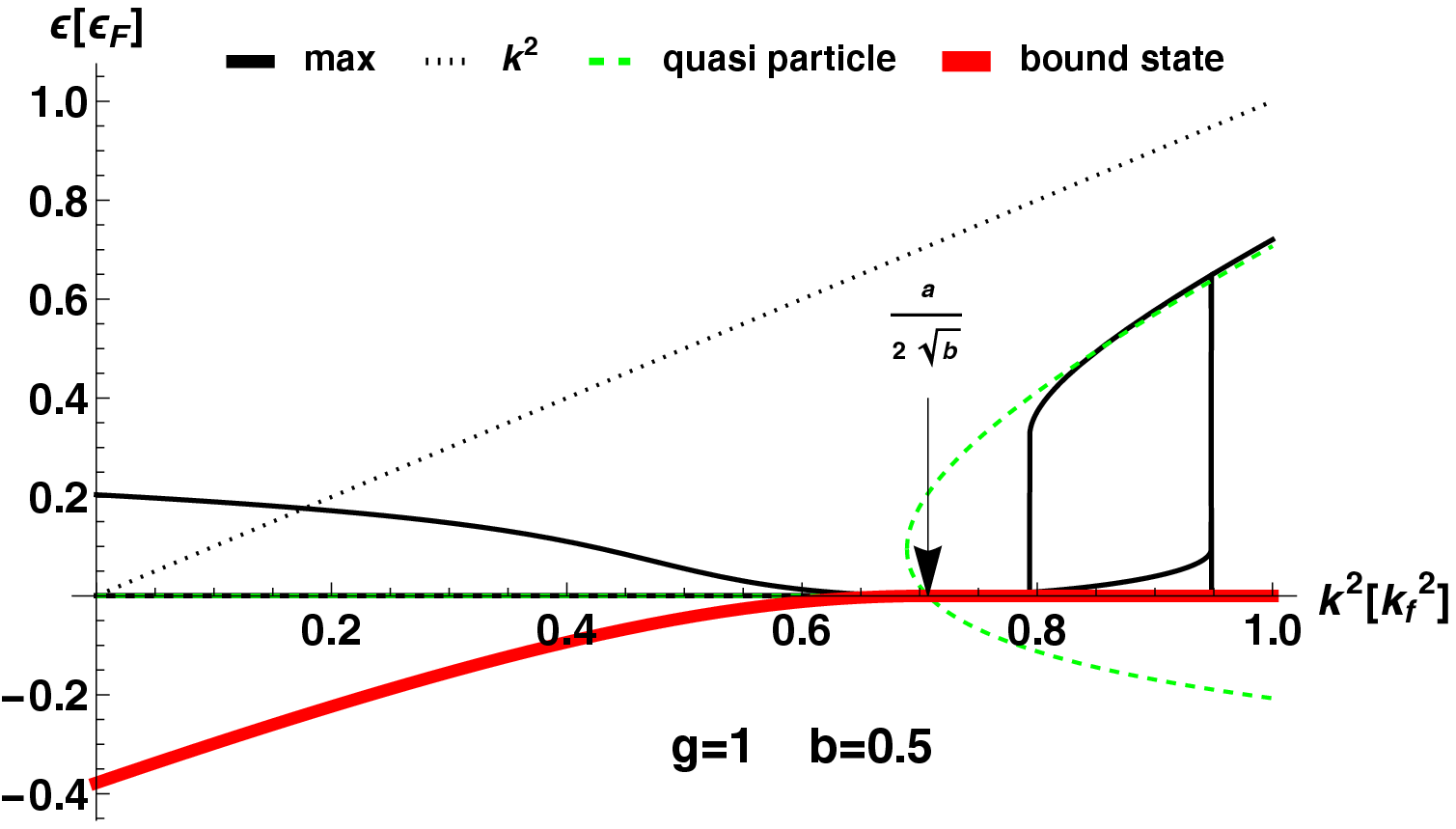}
\includegraphics[width=8.5cm]{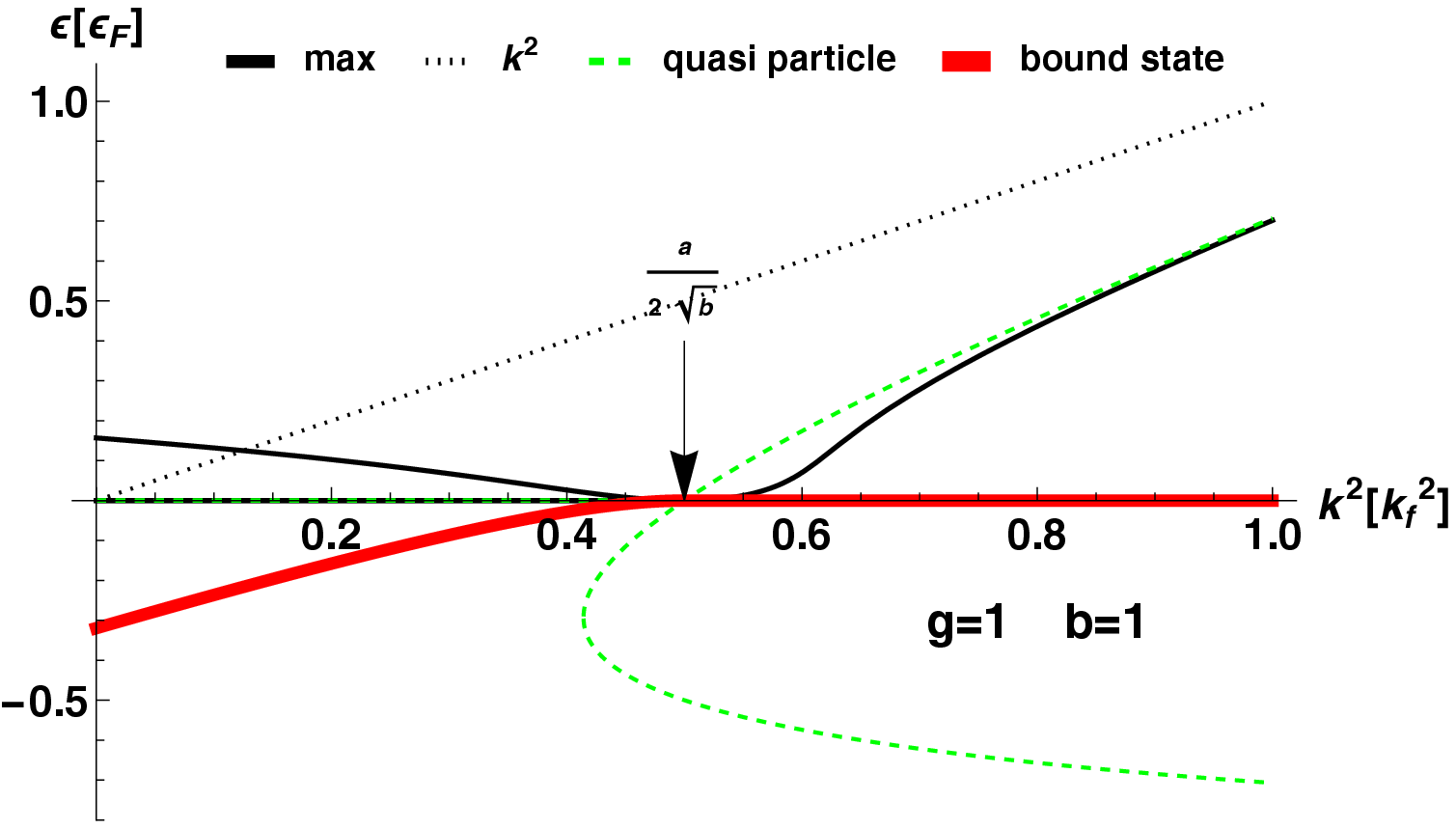}
\includegraphics[width=8.5cm]{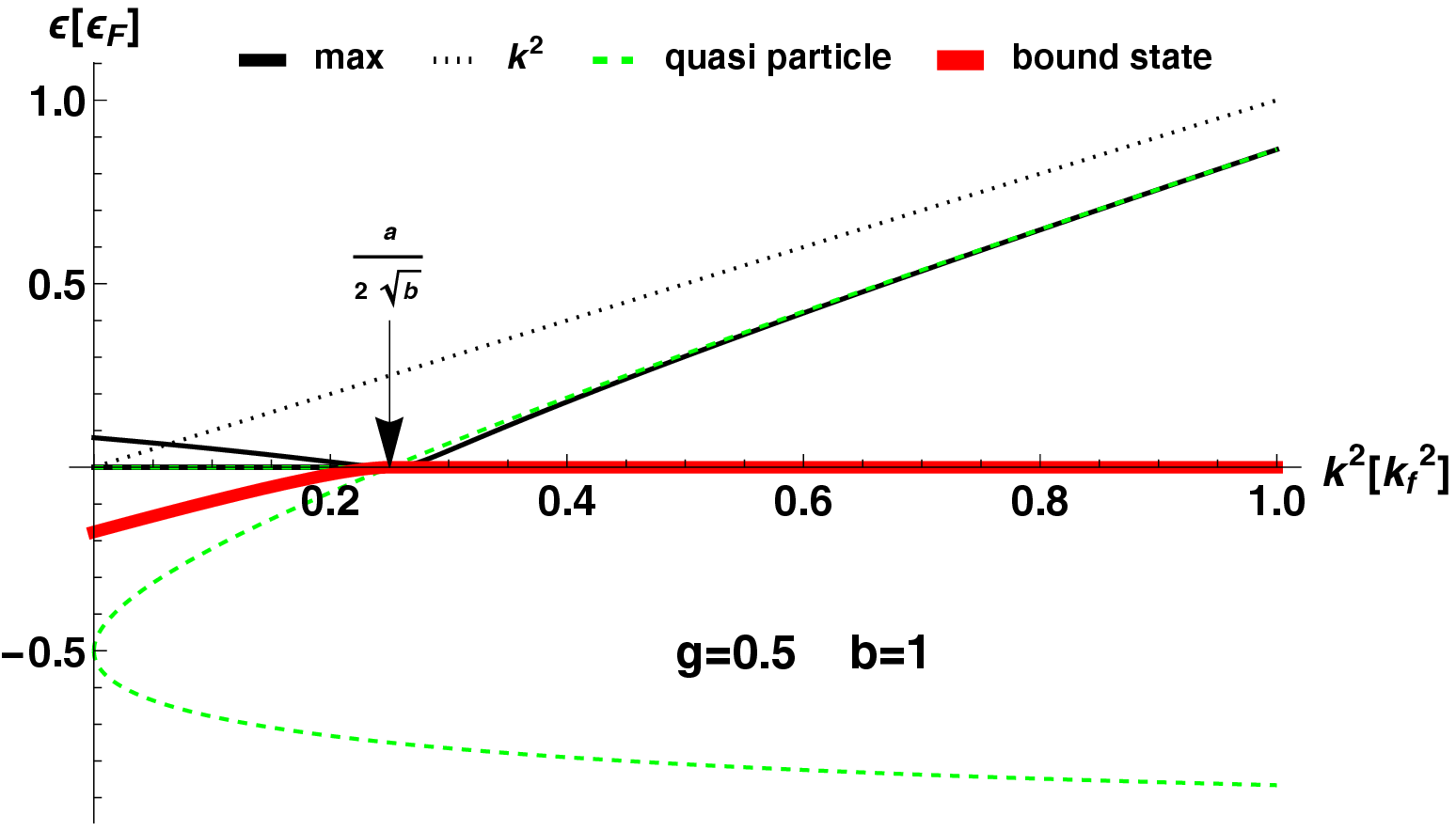}
\caption{The solution of dispersion (\ref{e12}) (dashed) together with the maximum of the spectral function (solid) and the negative bound-state pole (thick) for three different parameter sets.
\label{ep}}
\end{figure}

One sees that the bound-state pole vanishes at a certain momentum $k_0$ and resolves into an additional peak in the positive energy spectrum. This point can be found considering the quasiparticle dispersion. For $\omega>0$ we have two damped ($\gamma \ne 0$) solutions by
\be
\omega-\epsilon_k-\sigma(\omega)=0
\ee
as
\be
(\varepsilon_k)_{1,2}=\frac 1 2 (k^2-b\pm\sqrt{(k^2+b)^2-2 g\sqrt{b}}).
\label{e12}
\ee
For $\omega<0$ we have the bound-state pole $\varepsilon=-q_0^2$
as solution of
\be
q_0^3+\sqrt{b} q_0^2+ k^2q_0+k^2\sqrt{b}-{g\over 2}=0.
\label{q0T}
\ee
It is easy to see that possible extreme points of this polynomial are always at negative $q_0$. Since the polynomial increases with increasing $q_0$ we have only a real positive solution as a crossing of the polynomial with the $q_0$-axes if the value of the polynomial at $q_0=0$ is negative which translates into
\be
k^2<k_0^2={g\over 2\sqrt{b}}=r_s {n_i\over n_F}{2 s^3\over \pi^2}
\label{k0}
\ee
which determines the critical $k_0$ above which the bound-state pole vanishes and resolves into the positive dispersion. This value coincides with the value where the quasiparticle energy $\varepsilon$ of (\ref{e12}) crosses zero as illustrated in figure~\ref{ep}.

This turn-over of the bound-state pole from negative sharp values to positive damped values is accompanied by a sharp additional excitation as seen in figure~\ref{ep3D}. Such a localization in momentum space is typically for the onset of superconductivity as Bose condensation of pairs. One therefore suspects that around the momenta (\ref{k0}) one might have superconducting behaviour for a 1D wire interacting solely with impurities. We will find this interpretation supported by the temperature dependence of the conductivity in chapter~\ref{conduct}. Since this appears here with a finite momentum off the Fermi momentum we might have the situation similar to the FFLO state. Of course, any electron-electron scattering will probably smooth out this effect.

The quasiparticle dispersion (\ref{e12}) has only real solutions as long as
\be
k^2>k_c^2=\sqrt{2 g\sqrt{b}}-b.
\label{kc}
\ee 
One sees in figure~\ref{ep} that as soon as the most left point of this quasiparticle dispersion is turning from negative to positive values which is at $k^2= b$, the spectral function develops two maxima which can be seen in the top panel of figure~\ref{ep}. For higher momenta the quasiparticle dispersion and the points of maxima coincide. We have no minimal value of quasiparticle dispersion if according to (\ref{kc}) $2g<b^{3/2}$ which means as long as the impurity density does not exceed
\be
n_i<{1\over 8 a_B}
\label{nia}
\ee
which case is plotted in the bottom panel of figure~\ref{ep}.
It is quite remarkable that this limit does not depend on the density of electrons but only on the strength of interaction. 

As long as we have an impurity concentration higher than (\ref{nia})
the appearance of quasiparticle poles only above a finite momentum (energy) describes an energy gap. The same behaviour has been found as localized states in a pseudogap \cite{M68}. We will see indeed in chapter~\ref{conduct} that this leads to a minimum in the conductivity.  

\begin{figure}[h]
  \includegraphics[width=4.2cm]{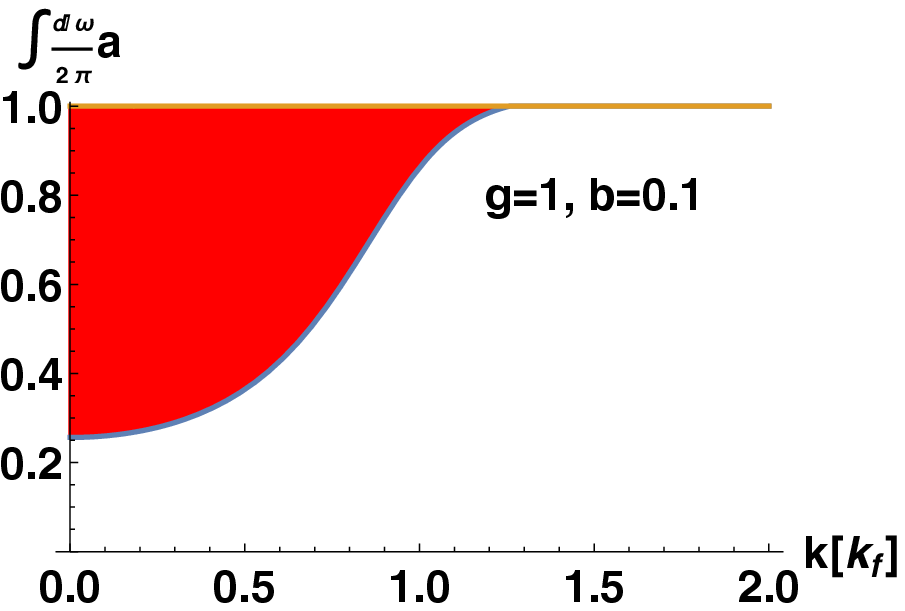}\includegraphics[width=4.2cm]{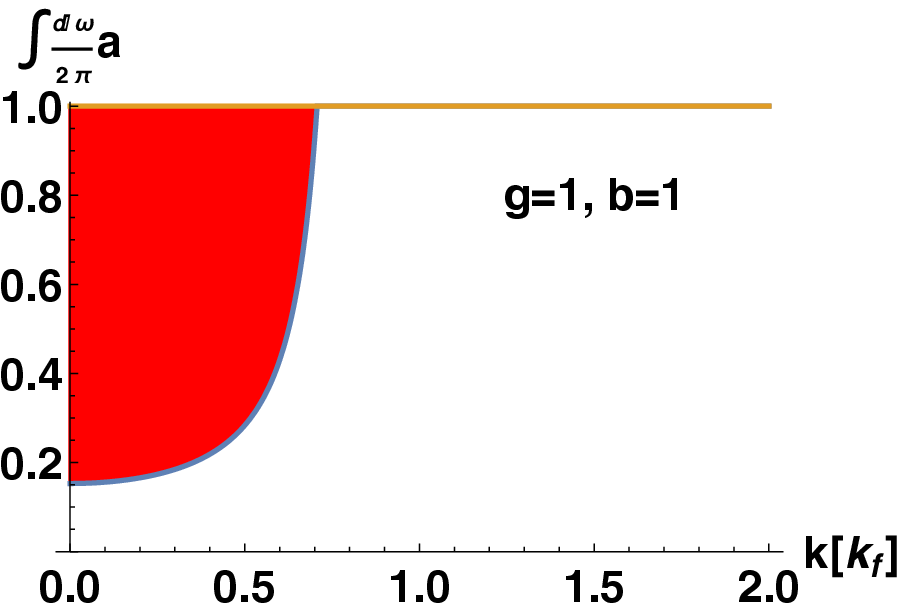}
\caption{The frequency sum rule for the impurity scattering model in T-matrix approximation  for different impurity couplings. The red area indicates the contribution of the bound-state pole (\ref{q0T}) at negative frequencies. 
\label{awt}}
\end{figure}

\begin{figure}[h]
\includegraphics[width=4.2cm]{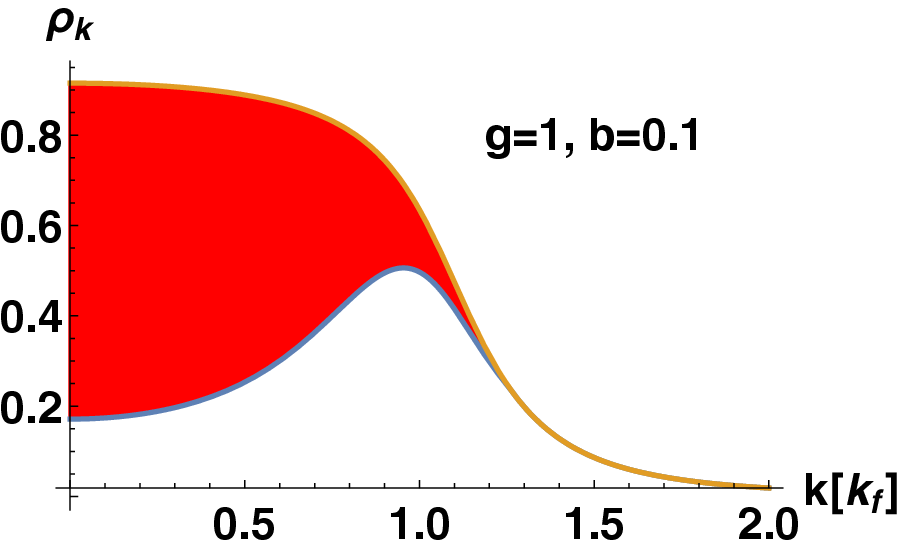}\includegraphics[width=4.2cm]{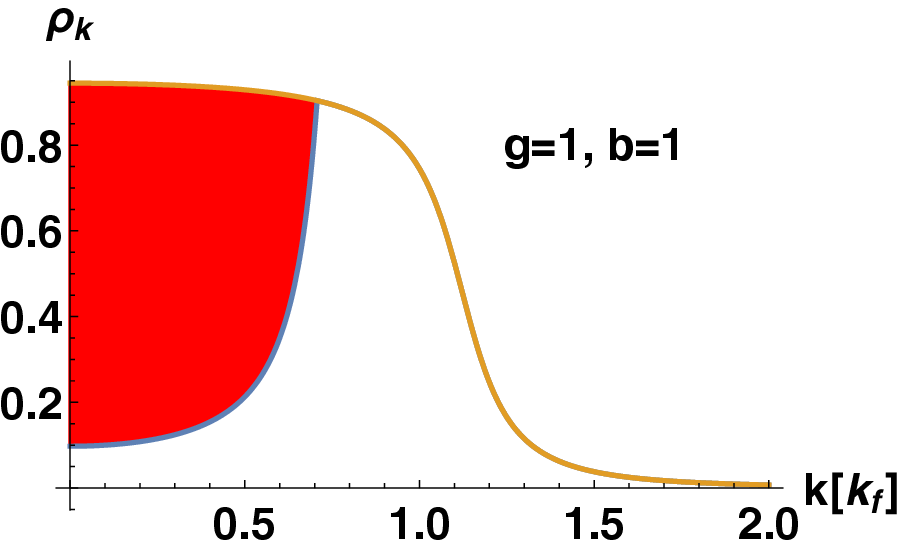}
\caption{The reduced density matrix (\ref{red2}) for the impurity scattering model in T-matrix approximation for two different impurity couplings corresponding to figure~\ref{awt}. The red area indicates the contribution of the bound-state pole at negative frequencies. 
\label{rhoT}}
\end{figure}

To complete the discussion we illustrate again the importance of the bound-state pole to the sum rule (\ref{sumr}) in figure~\ref{awt}.
The corresponding reduced density matrix (\ref{red2}) are plotted in figure~\ref{rhoT} which again illuminates the influence of the bound-state pole at negative frequencies. Compared to the Born approximation in figure~\ref{aw} and figure~\ref{rho}
 the second parameter $b$ of (\ref{gb}) creates different shapes of the momentum dependence.

\subsubsection{Pad\'e approximation}

The Pad\'e approximation (\ref{t1}) reads now for the impurity model in T-matrix approximation (\ref{tmat}) 
\ba
&\rho_{1\!-\!{\rm f}}=\Theta(1-k^2)
\nonumber\\&-
\!\left (\!\!
\Theta(1\!-\!k^2)\!\!\int\limits_1^\infty 
\!\!-
\Theta(k^2\!-\!1)\!\!\int\limits^1_{0} \!
\right )\!\!{dq \over 2 \pi}\gamma(q^2)
\partial_q{q^2\!-\!\varepsilon_k\over (q^2\!-\!\varepsilon_k)^2\!+\!{\gamma(\varepsilon_k)^2
    \over 4}}
\label{r1fT}
\end{align}
with an elementary integral due to (\ref{tmat}). The second part (\ref{t2}) becomes
\ba
\rho_{\rm f}=&{1\over 1+{g\over 4 q_0(\sqrt{b}+q_0)^2}}+\frac{1}{ \pi} {\rm arctan}{2\over \gamma}
\nonumber\\&-{(k^2-1) \over 2 \pi} \left [ {\gamma\over 
 (1-\varepsilon_k)^2+{\gamma^2\over 4}}-{\gamma\over 
 \varepsilon_k^2+{\gamma^2\over 4}}\right ]
\label{rfT}
\end{align}
where $\gamma=\gamma(\epsilon_F)$ and all energies are in terms of $\epsilon_F$. We had to break the integration over frequency again into parts larger and smaller zero where the latter leads to the bound-state pole contribution as the first part in $\rho_{\rm f}$.

\begin{figure}[h]
\includegraphics[width=8.5cm]{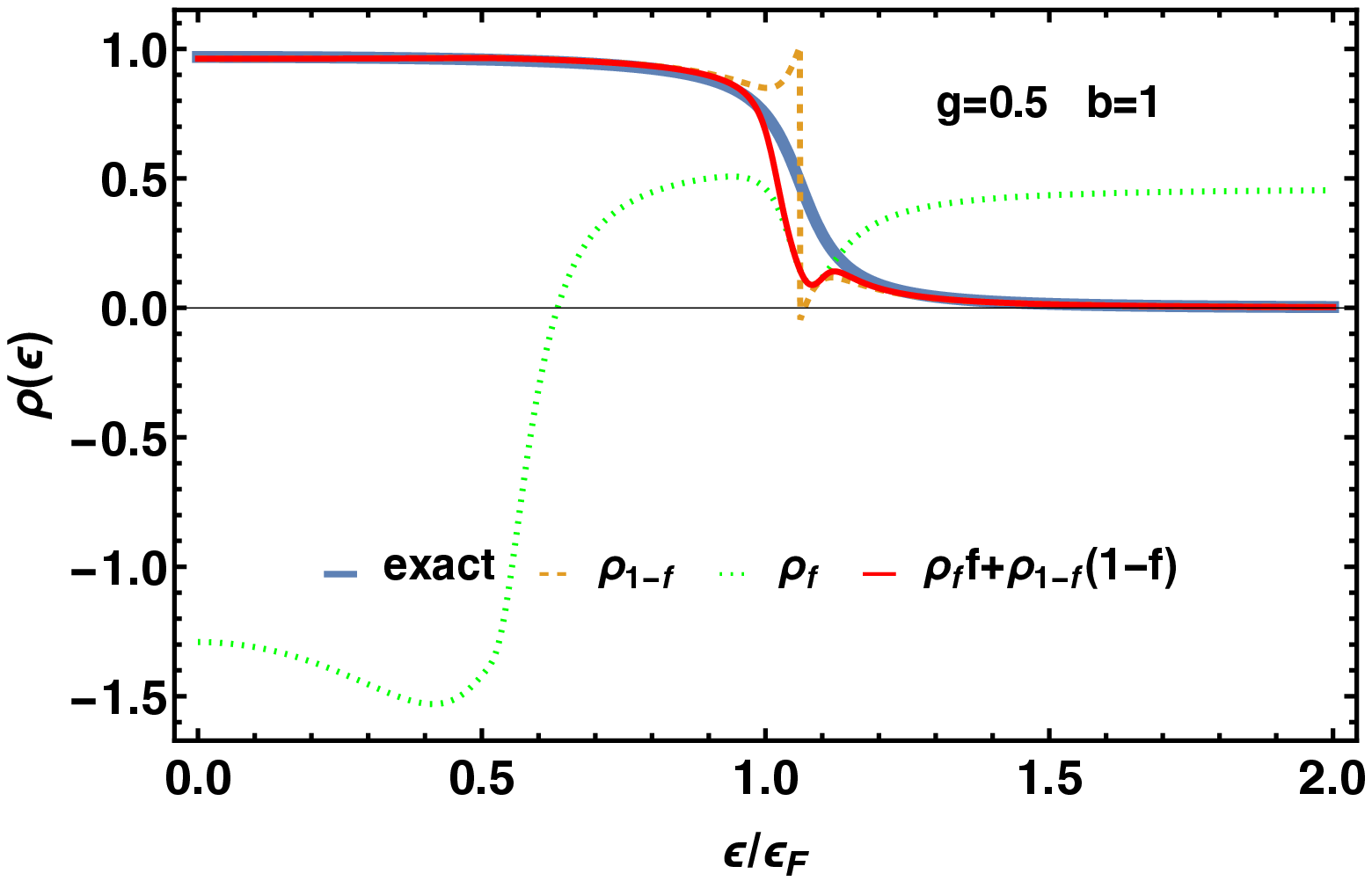}
\includegraphics[width=8.5cm]{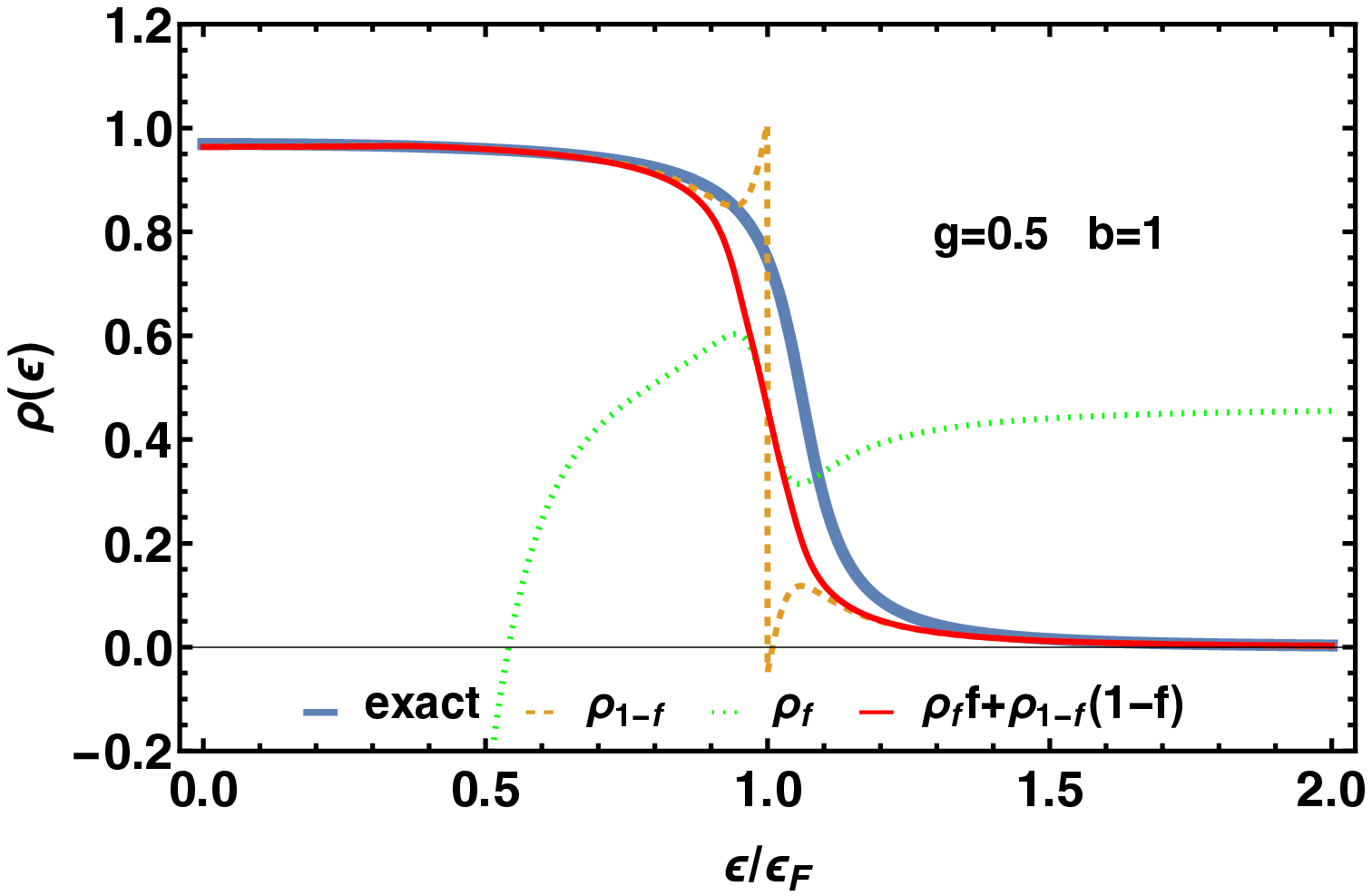}
\includegraphics[width=8.5cm]{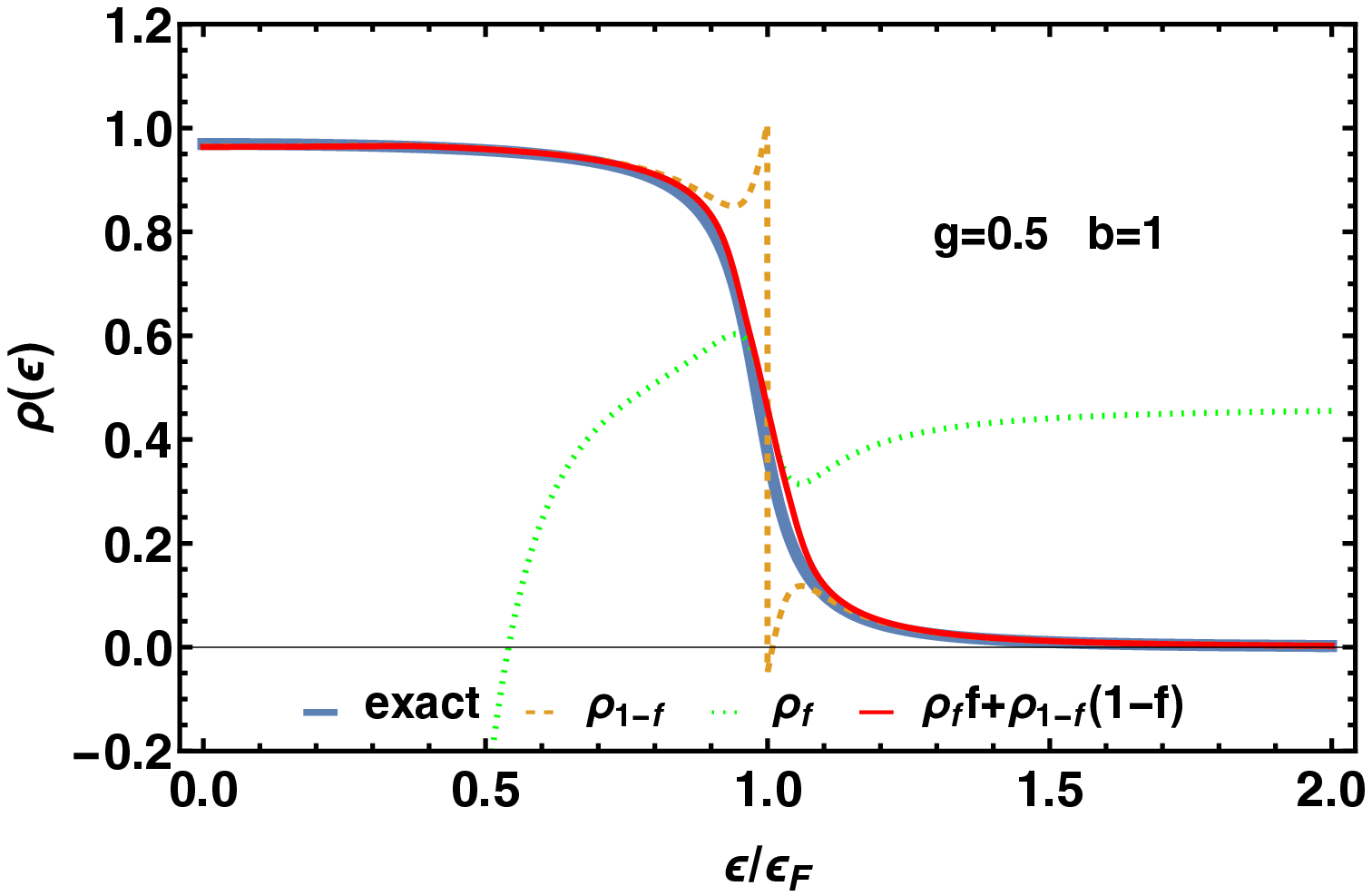}
\caption{The reduced density matrix in Pad\'e approximation (thin red) compared with (\ref{red2}) (thick blue) and both contributions entering the Pad\'e approximations, $\rho_{\rm 1-f}$ of (\ref{r1fT})  (dashed) and $\rho_{\rm f}$ of (\ref{rfT}) (dotted). Above: without selfconsistent quasiparticle and without correlated density, middle: with selfconsistent quasiparticle and without correlated density, below: with selfconsistent quasiparticle and correlated density. 
\label{rhoTed}}
\end{figure}

If we compare this Pad\'e approximation with the reduced density matrix from the spectral function (\ref{red2}) in figure~\ref{rhoTed} (above) we see a strong deviation. It is instructive to reveal the missing parts in two steps. 

\paragraph{Selfconsistent quasiparticles}

First we have used the quasiparticle energy at values of the free energy $\varepsilon_k=\varepsilon(k^2)$. A selfconsistent solution of the quasiparticle dispersion (\ref{pole}) would require a successive iteration which means $\varepsilon_k=\varepsilon(k^2+\delta)$. We can determine this required shift $\delta$ by demanding that the Fermi energy should be reproduced $\epsilon_F=\varepsilon(k_F^2+\delta)$. For the quasiparticle approximation (\ref{e12}) one finds explicitly $\delta=-\sigma(\epsilon_F)$. Including this shift leads to a smoothing of the curve parallel to the spectral expression (\ref{red2}) as seen in the middle panel of figure~\ref{rhoTed}. 

\paragraph{Correlated density}

Second, we recognize a parallel shift between the spectral result (\ref{red2}) and the Pad\'e approximation in the middle panel of figure~\ref{rhoTed}. This comes from the fact that the T-matrix approximation leads to time nonlocality in the scattering process given by the delay time $\Delta_t=\partial_\omega \Phi$ between in- and outgoing scattering events with the phase of the T-matrix $T^R=|T|{\rm e}^{i\Phi}$. As result a correlated density $n_c$ appears besides the free (quasiparticle) density adding to the total density $n=n_F+n_c$. It describes how much particles are in the correlated state not available for the Fermi momentum. Consequently it diminishes the Fermi momentum \cite{SLMa96}
\begin{equation}
k_F={ \pi\hbar\over s}n_F={ \pi\hbar\over s} {n\over 1+{n_c\over n}}.
\label{tc2c}
\end{equation}
This shift due to the time delay clearly violates the Luttinger theorem stating that the Fermi momentum is exclusively determined by the free (quasiparticle) density $n_F$. Since the Luttinger theorem is valid only for Fermi liquids, one should not be puzzled since we do have a non-Fermi liquid.  Collision delays describe the correlations beyond Fermi liquids. Other examples of non-Fermi liquids are the correlated density for systems with short-living bound states (resonances) or pairing. 

Before presenting the explicit form we see already in the bottom panel of figure~\ref{rhoTed} that the missing agreement between the Pad\'e T-matrix approximation and the spectral expression of the reduced density (\ref{red2}) is completed by this time delay respectively correlated density. The comparison of the momentum distribution with results of other methods, e.q. by variation technique with figure 4 in \cite{SZ19}, can be done but is not sensitive enough to distinguish the subtle differences.

\section{Time delay and correlated density}

\subsection{Time-delay and correlated density}
\begin{figure}[h]
\includegraphics[width=8.5cm]{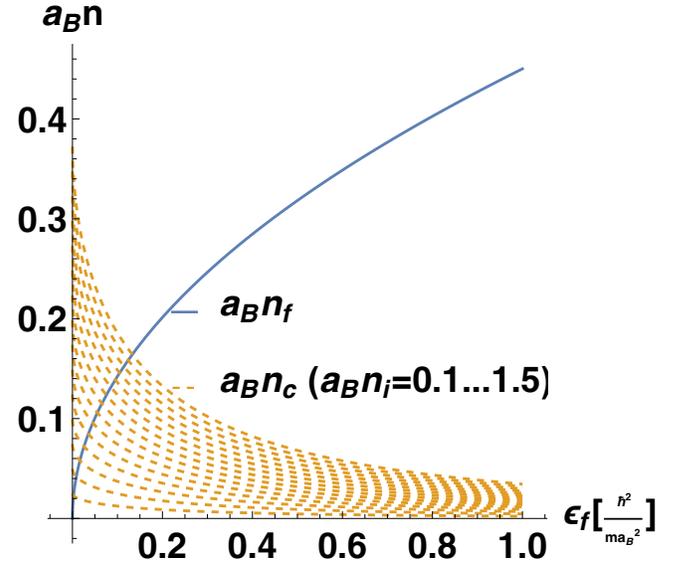}
\caption{The free and correlated density vs. Fermi energy for increasing impurity densities $a_B n_i=0.1-1.5$ (from bottom to top). 
\label{nf_n_ef}}
\end{figure}

As soon as the vertex, the T-matrix $T=|T|\exp{i\Phi}$, has a frequency-dependent phase $\phi(\omega)$ there appears a finite duration of the collision process. The first-order effect can be  
expressed (see Eq. 68 in \cite{MLS00})  as a time delay $\Delta_t=\partial_\omega \Phi$ correction to the selfenergy (\ref{ssmall})
\be
&&\Delta \sigma^<(\omega,k)=s\sum\limits_{p,q}|T^R(\omega\!+\!\varepsilon_p^i,k,p)|^2 \Delta_t \nonumber\\&&\times 2\pi \delta(\omega\!+\!\varepsilon_p^i\!-\!\varepsilon_{k\!+\!q}\!-\!\varepsilon_{p\!-\!q}^i)(1-n_p^i){\partial\over \partial t} n_{k+q}n_{p-q}^i.
\label{ssmall1}
\ee
We express now 
\be
|T^R|^2\Delta_t={\rm Im} T^A \partial_\omega T^R={\hbar^2\over k_F^2a_B^2} {\sqrt{b\omega}\over 2(\omega+b)^2}
\label{delT}
\ee
and use the impurity limit $1-n_p\approx 1$, $n_{p-q}\approx n_p$ to find the additional contribution to the collision integral $I=(1-n_k)\sigma^<(\varepsilon_k)-n_k\sigma^>(\varepsilon_k)$ to be
\be
\Delta I=&{\partial\over \partial t} n_i s \sum\limits_{q}{\rm Im} T^A \partial_\omega T^R 2\pi \delta(\varepsilon_k\!-\!\varepsilon_{k\!+\!q}) n_{k+q}.
\label{ssmall2}
\ee
This means that in the balance equation for the density we obtain with (\ref{delT}) an additional correlated density contribution
\be
n_c&=&-n_i s \sum\limits_{q}{\rm Im} T^A \partial_\omega T^R 2\pi \delta(\varepsilon_k\!-\!\varepsilon_{k\!+\!q}) n_{k+q}
\nonumber\\
&=&-n_i{g\over 8 s}\int\limits_{-1}^1 dq {\sqrt{b}\over (q^2+b)^2}.
\label{nc1}
\ee
For Born approximation $b\to 0$ we do not have a phase and the collision delay $\sim \sqrt{b}$ is zero. Therefore the integral (\ref{nc1}) should vanish. This is the case if we expand the integrand into a Taylor series and get
\be
\int\limits_{-1}^1 dq {\sqrt{b}\over (q^2+b)^2}=-\frac 2 3 \sqrt{b}+\frac 4 5 b^{3/2}+....
\ee
Unfortunately the integral does not smoothly converge which means the integration and expansion cannot be interchanged. In fact one gets formally
\be
\int\limits_{-1}^1 dq {\sqrt{b}\over (q^2+b)^2}&=&{1\over \sqrt{b}(1+b)}+\frac 1 b {\rm arctan}{1\over \sqrt{b}}
\nonumber\\
&=&{\pi\over 2 b}-\frac 2 3 \sqrt{b}+\frac 4 5 b^{3/2}+....
\ee
which compared to (\ref{nc1}) shows that the term $\pi/2b$ appears to be wrong in the sense of perturbation expansion of the integrand which we demand as limit of T-matrix towards Born approximation. Therefore we subtract this term which means we first expand and then integrate and then resume again to get finally
\be
n_c&=&-n_i{g\over 8 s}\left [
{1\over \sqrt{b}(1+b)}-\frac 1 b {\rm arctan}{\sqrt{b}}\right ].
\label{nc}
\ee
This correlated density adds to the free density for total density $n=n_F+n_c$ and accordingly changes the Fermi momentum since the latter one is determined with respect to the free density (\ref{tc2c}). In figure~\ref{nf_n_ef} the free density increases with the square of Fermi energy while the correlated density decreases with increasing Fermi energy dependent on the impurity density. This different dependence will lead to localization effects in the conductivity as we will see in chapter~\ref{conduct}.

\subsection{Tan contact and large momentum behaviour}

\begin{figure}[h]
\includegraphics[width=8.5cm]{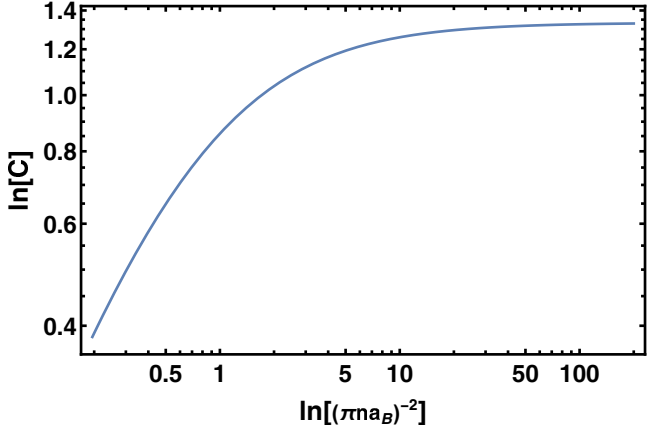}
\caption{The large momentum expansion of the reduced density $\rho(k)\approx C(n \hbar)^4/k^4$ versus coupling constant (\ref{gb}).
\label{C}}
\end{figure}

The expansion at large momentum in the Pad\'e approximation only the term $\rho_{\rm 1-f}$ contributes as it is in the extended quasiparticle approximation. From (\ref{e12}) and  (\ref{r1fT}) we obtain
\be
\lim\limits_{k\to\infty}\rho(k)={C n^4\hbar^4\over k^4}
\ee
with the Tan contact
\be
C&=&{gk_F^4\over \pi n_F^4\hbar^4}\left [\sqrt{b}\,{\rm arccot}\sqrt{b}-1\right ]
\nonumber\\
&=& {4 n_i\over n_F} \pi^2 b\left [\sqrt{b}\,{\rm arccot}\sqrt{b}-1\right ]
\label{CC}
\ee
independent of the pole correction $\delta$.
The values for the e-e model of contact interaction is compared in \cite{Ram17}. In figure~\ref{C} we plot this contact (\ref{CC}) versus the coupling constant $b$ of (\ref{gb}). It is qualitatively very similar to the e-e interaction considered in \cite{OD03} where it had been conjectured its universal behaviour.
This is quantitatively supported when comparing with the result of Eq. 10 in \cite{SZ19} where an impurity immersed in Fermions forms a polaron. Reformulating  
for repulsive interaction their result reads \footnote{The strange factor $2$ in the subtraction instead of $1$ with our result might be a misprint.
}
\be
C&=& {1\over 2 a_B n_F} \pi^2 b\left [\sqrt{b}\,{\rm arccot}\sqrt{b}-2\right ]
\label{CC1}
\ee
and compared to (\ref{CC}) we see that we have to replace in our result the impurity density simply by $n_i=1/8a_B$ to reproduce the result of \cite{SZ19}. Remarkably, this is just coinciding with the limit of (\ref{nia}) where the spectral function changes from one to two maxima and forming a gap. 

\section{Nonlocal corrections to the conductivity\label{conduct}}

For any transport property we have to consider momentum integrals over the reduced density abbreviating $x=2 (\epsilon_F\!-\!\varepsilon_k)/\gamma(\epsilon_F,k)$
\be
\rho^{\rm Pade}(k)=\rho^{\rm EQP}(k)+f(x) [\rho_{\rm f}(k)-\rho_{\rm 1-f}(k)]
\label{excess}
\ee
which presents the difference to the extended quasiparticle part $\rho^{\rm EQP}=\rho_{\rm 1-f}$. In appendix~\ref{cancel} we show that the additional part is negligible under integration for most regular cases of selfenergy. This means though we have to use two-term Pad\'e approximation to reproduce the spectral function and the reduced density matrix at the Fermi energy correctly, only the part of the extended quasiparticle part matters under integrations. In other words the divergence of the extended quasiparticle distribution function at the Fermi energy cancels under integration.

The impurity scattering including averaged T-matrix approximation contains the average about impurity configurations and therefore reinforce spatial invariance. We write the kinetic equation for homogeneous systems in an electric field $E$
\ba
\p t n_k \!-\!e E\p k n_k\!=\!&n_is\!\!\!\int\limits_{-\infty}^\infty\!\!{d q\over \hbar^2}|T(\varepsilon_{k\!+\!q})|^2\delta(\varepsilon_{k\!+\!q}\!-\!\varepsilon_{k})[n_{k\!+\!q}\!-\!n_k]
\nonumber\\&+\Delta I
\label{kinetic}
\end{align}
with the nonlocal corrections (\ref{ssmall2}). When linearized with respect to the electric field
\be
n(k,t)=n_0(|k|)+\tau(|k|) e E \p k n_0(|k|)
\ee
one obtains the relaxation time
\be
\tau(|k|)&=&|k| {h^2\over 2 m n_i s T^2(\epsilon_k)}={\hbar\over g \epsilon_F}\sqrt{\epsilon_k\over \epsilon_F}\left ( 1+{b\over \epsilon_k/\epsilon_F}\right )
\nonumber\\
\tau(k_F)&=&
{\hbar \over \epsilon_F}{\pi\over 2 s^2} {n_F\over n_i} {1+b\over b}.
\ee
The energy dependence of this relaxation rate starts at zero showing a maximum and vanishes for large energies. Coulomb interactions between the electrons just lead to the opposite behaviour \cite{RM21}.

\begin{figure}[h]
\includegraphics[width=4.2cm]{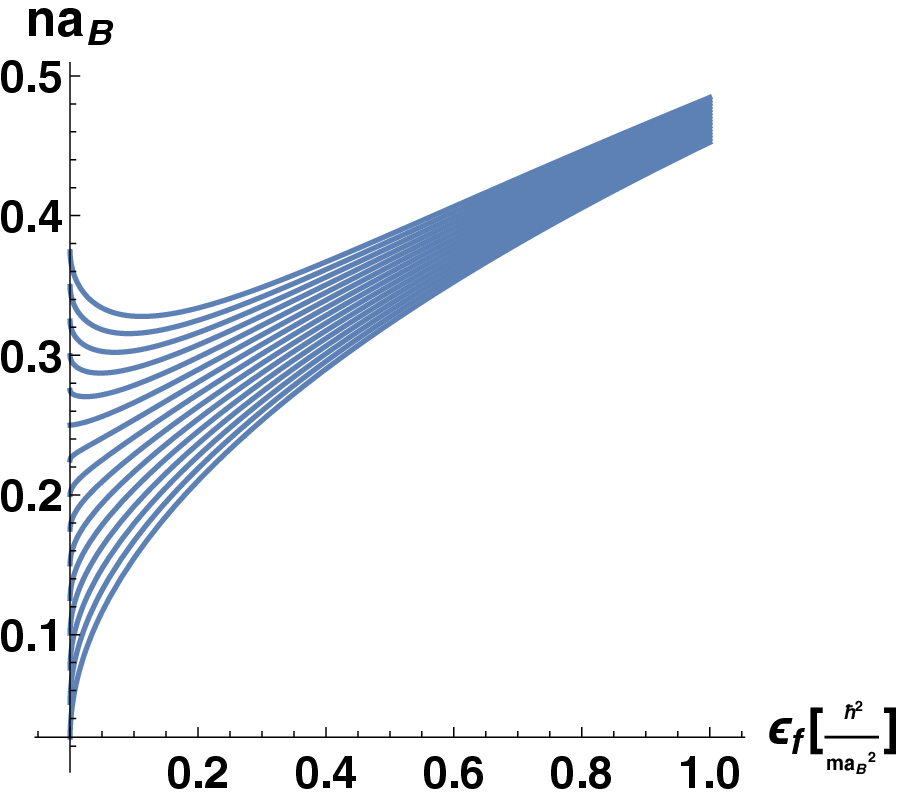}
\includegraphics[width=4.2cm]{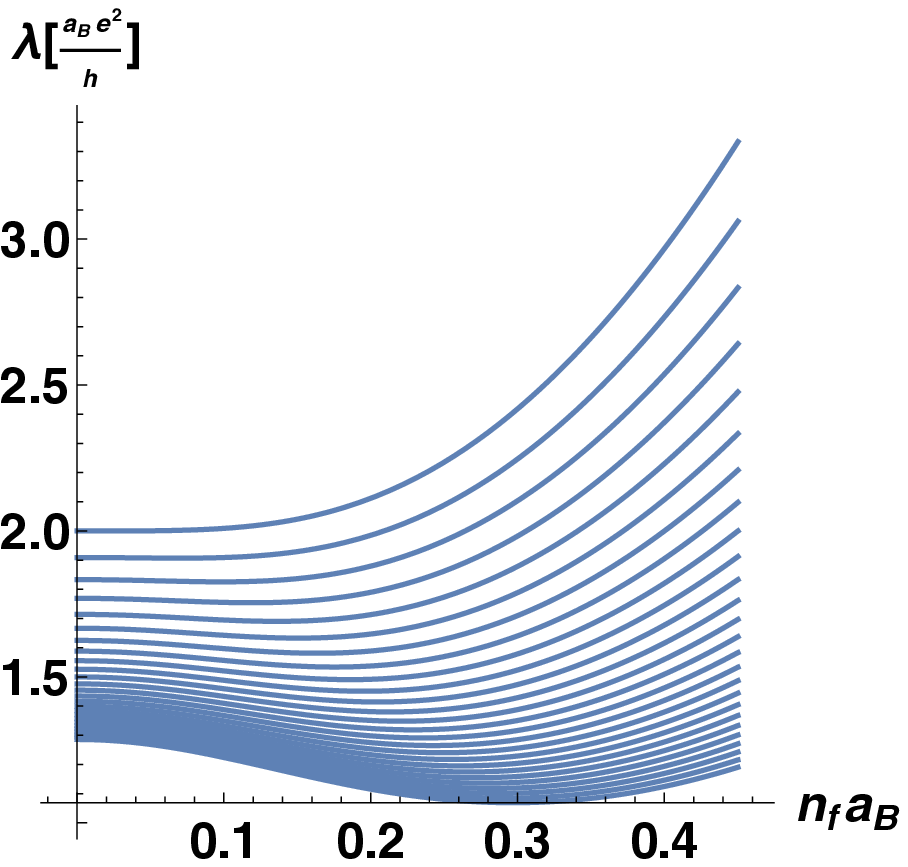}
\caption{The total density  versus Fermi energy  (left) and conductivity versus free density (right) for zero temperature (\ref{sT0})
with $n_ia_B=0.1-1.5$ from bottom to top (left) and $n_ia_B=1-3.5$ from top to bottom (right). 
\label{sigmaT}}
\end{figure}

By integrating the kinetic equation (\ref{kinetic}) with momentum weight, the balance for the current becomes
\be
\p t (j+j_c)=-e E n_F
\ee
and leads besides the standard current density
\be
j=-e E\int{dk\over 2 \pi \hbar} {k\over m} f=\lambda_0 E
\ee
also to a current density due to two-particle correlations
\be
j_c\!&=&\!e n_is\!\!
\int\limits_{-\infty}^\infty\!\!{d q d k\over 2\pi \hbar^2}{k\!+\!q\over m}\, {\rm Im} T^RT^A\, \delta(\varepsilon_{k+q}\!-\!\varepsilon_{k})n_{k+q}
\nonumber\\&=&\lambda_c E.
\ee
The corresponding conductivities are
\be
\lambda_0\!&=&\!{n_F e^2\tau(p_F)\over m}\!=\!{e^2\over n_i \hbar} {\pi \over 2 s^4 r_s^2}(1\!+\!b)={e^2\over  2 \pi \hbar n_i} {1\!+\!b\over b}\nonumber\\
\lambda_c&=&\lambda_0 {s\over \pi} {n_i\over n_F} {b^{3/2}\over (1+b)^2}
\label{sT0}
\ee
where we used (\ref{gb}).
The results for the density (\ref{nc}) and total conductivity $\lambda=\lambda_0+\lambda_c$ with (\ref{sT0}) can be seen in figure~\ref{sigmaT}. One recognizes that both the total density and the total conductivity lead to a development of a minimum at small electron/impurity density ratios due to the different behaviour of the free and correlated density as seen in figure~\ref{nf_n_ef}.

\begin{figure}[h]
\includegraphics[width=8.5cm]{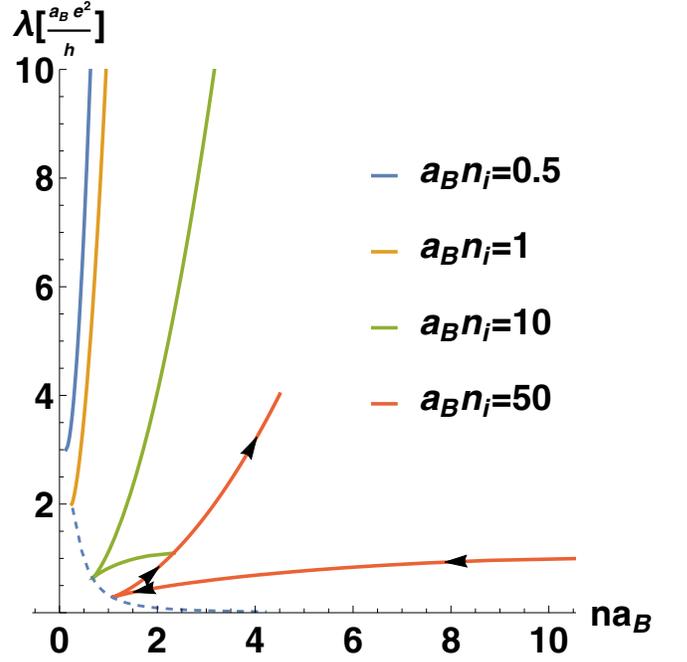}
\caption{The conductivity versus total density for different impurity densities and the Fermi energy as parameter denoted by the arrows.  For $n_ia_B>1$ the curves start at zero Fermi energy at $(n_ia_B/2, 1+1/n_ia_B)$ and develop a minimum at the energy $\epsilon^{\rm min}=(\sqrt{n_ia_B}-1)/2$ and $\lambda^{\rm min}=e^2a_B /\pi \hbar\sqrt{n_i a_B}$ when the Fermi energy is increasing. Bellow $n_ia_B<1$ no such minima appears. The localization of the minima are plotted as dotted line according to (\ref{limit}). 
\label{sigma_n}}
\end{figure}

These minima lead now to a localization effect in the conductivity when plotted versus total number of particles as done in figure~\ref{sigma_n}. The minima appear as long as $n_ia_B\ge 1$ which all lay on the curve
\be
n_i^{\rm min}a_B\!&\!=\!&\!\!{n_i a_B\over 2 \pi} {\rm arccot}{\sqrt{\!\sqrt{n_i a_B}\!-\!1}}\nonumber\\&&-\left (\!{\sqrt{n_i a_B}\over 2 \pi}\!-\!{1\over \pi}\!\right)\sqrt{\!\sqrt{n_i a_B}\!-\!1}\nonumber\\
\lambda^{\rm min} &=&{e^2 a_B \over \pi \hbar \sqrt{n_i a_B}}
\label {limit}
\ee
easily seen from (\ref{nc}) and (\ref{sT0}).

\subsection{Temperature dependence}
\begin{figure}[h]
\includegraphics[width=8.5cm]{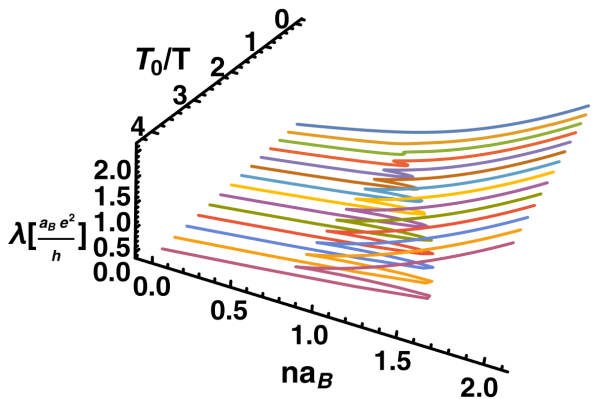}
\caption{The conductivity versus total density and inverse temperature for an impurity density $n_ia_B=20$. 
\label{sigma_n_T}}
\end{figure}

Since the expressions for the conductivity are linear in the distribution functions we can calculate the finite temperature $T$ expressions observing the relation \cite{Mal78}
\be
n_k=\int\limits_0^\infty d\bar \mu {\Theta (\bar \mu-\varepsilon_k)\over 4 T {\rm cosh}^2{\mu-\bar \mu \over 2 T}}
\ee
to be applied to the zero-temperature expressions (\ref{sT0})
\be
\lambda(0)\!=\!\lambda_0(0)\!+\!\lambda_c(0)\!=\!{e^2\over 2 \pi n_i \hbar}\!\left (\!
1\!+\!{k_F^2 a_B^2\over \hbar^2}\!+\!{n_i a_B\over 1\!+\!{k_F^2 a_B^2\over \hbar^2}}\!
\right )
\ee
where the last term is the correlated conductivity.
The integral about the normal conductivity can be done analytically to yield
\be
\lambda_0(T)={e^2\over 2\pi n_i \hbar}\left [{z\over 1+z}+2{T\over T_0} \ln(1+z)\right ]
\ee
with the temperature scale and fugacity
\be
T_0&=&{\hbar^2\over m a_B^2}\nonumber\\
z&=&{\rm e}^{\mu\over kT}
\ee
where the last one determines the free density
\be
n_F a_B= -\sqrt{2 T\over \pi T_0}Li_{1/2}(z)
\ee
by the poly-logarithm $Li_n(z)=\sum_{k=1}^\infty z^k/k^n$.
The correlated conductivity remains as integral
\be
\lambda_c(T)={e^2 a_B\over 4 \pi \hbar}\int\limits_0^\infty {d y\over \left (1+{2T\over T_0} y\right )[1+\cosh(y-\ln z)]}.
\ee
It is remarkable that this expression becomes independent on the impurity density and is only determined by the interaction strength coded in the equivalent Bohr radius $a_B$ according to $V_0=\hbar^2/ma_B$ and by the density of electrons.
The effect of temperature is basically to diminish the area of localization seen as shrinking hysteresis in figure~\ref{sigma_n_T}.

In the next figure~\ref{sigma_mu_T} we plot the conductivity for various temperatures versus the chemical potential. One sees that for lower temperatures it develops a peaked structure and a minimum. This is the behaviour observed in figure 3 of \cite{PAC16}. Therefore we interpret this as the onset of a possible superconducting behaviour triggered by impurities. Without the two-particle correlations expressed by the delay time or correlated density this effect would be absent.
 
\begin{figure}[h]
\includegraphics[width=8.5cm]{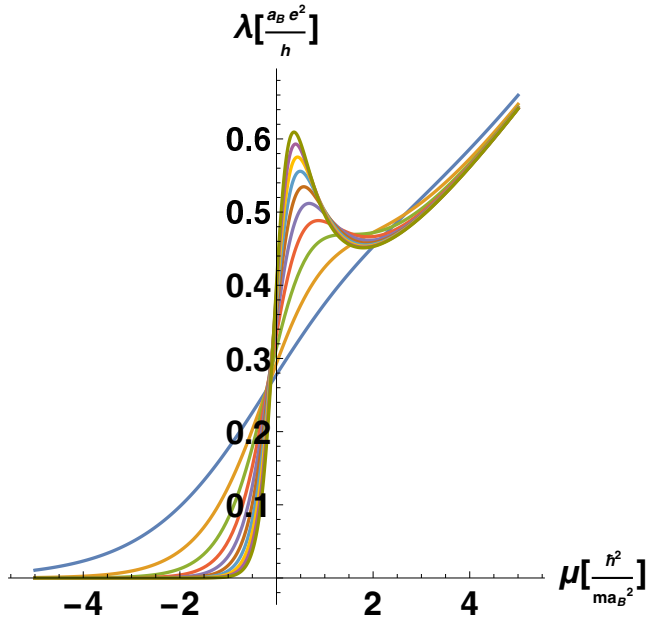}
\caption{The conductivity versus chemical potential for temperatures $T=0,3...2.5 \hbar^2/ma_B^2$ (from top to bottom of peak) and an impurity density $n_ia_B=20$. 
\label{sigma_mu_T}}
\end{figure}

\section{Summary}

A two-fold expansion of the spectral function in interacting quantum systems is presented. One expansion with respect to small scattering rates is the known extended quasiparticle picture. This expansion fails in non-Fermi systems like one-dimensional Fermi gases since it diverges at the Fermi energy. Therefore a second expansion is suggested around the Fermi energy. Both expansions can be interpolated by a Pad\'e expression. The advantage is that independent on the interpolating function the first two energy-weighted sum rules are completed as it is the case in the extended quasiparticle picture. The validity of the approach is the same as the perturbation theory where the extended quasiparticle picture is based on. With the help of the presented Pad\'e regularization the divergence of the reduced density in perturbation theory is cured at the Fermi energy independent on the used selfenergy approximation.

For models of electron-impurity scattering the quality of this Pad\'e approximation is demonstrated to reproduce the spectral function and reduced density. In this way the divergence of the reduced density in extended quasiparticle approximation at the Fermi energy is corrected to a finite value. It is shown that the divergence of the extended quasiparticle approximation at the Fermi energy cancels under integration. Consequently, in order to describe the reduced density and spectral function a two-term Pad\'e interpolation is necessary while for transport processes the extended quasiparticle approximation alone is sufficient. This extends the validity of extended quasiparticle approximation beyond the Fermi-liquids.

For the T-matrix approximation one obtains delay times resulting into a correlated density which describes how much particles remain in a correlated state. Further consequences are an additional contribution to the conductivity which results into two effects. At certain impurity densities localization appears and a minimum of conductivity develops for small temperatures dependent on the impurity concentration which is interpreted as a possible onset of superconductivity. 
The Tan contact shows the same universal behaviour as found in e-e scattering when we replace the impurity concentration by the value where the gap in the spectral function vanishes.

The Pad\'e expansion of the spectral function is derived. 
Here it is illustrate on a simple exploratory example on electron-impurity scattering. In a forthcoming paper it will be demonstrated that it works equally well for electron-electron correlations. The expectation is that it might be fruitfully applicable also for other non-Fermi systems.

\acknowledgements
The motivating discussions with Kare N. Pathak and Vinod Ashokan are gratefully acknowledged as well as the support from DFG-project
MO 621/28-1.

\appendix
\section{Derivation of Euler expansion\label{Eulerderv}}

The expansion of $arctan$ function found by Euler in 1755 reads
\ba
&\arctan x=\int\limits_0^1 dt {x\over 1+x^2 t^2}
=\int\limits_0^{\pi/2} du {x \sin{u}\over 1+x^2 \cos^2{u}}
\nonumber\\
&
=\int\limits_0^{\pi/2} du {x \sin{u}\over 1+x^2}{1\over 1- {x^2\sin^2{u}\over 1+x^2}}
=
\int\limits_0^{\pi/2} du \sum\limits_{n=0}^\infty {x^{2 n+1} \sin^{2 n+1}{u}\over (1+x^2)^{n+1}}
\nonumber\\&
=
\sum\limits_{n=0}^\infty\left (\int\limits_0^{\pi/2} du \sin^{2 n+1}{u}\right ) {x^{2 n+1} \over (1+x^2)^{n+1}}
\nonumber\\&
=
\sum\limits_{n=0}^\infty {2^{2n}(n!)^2\over (2 n+1)!} {x^{2 n+1}\over (1+x^2)^{n+1}}
={x\over 1 +x^2}+\frac  2 3 {x^3\over (1+x^2)^{2}}+...
\end{align}
with obvious substitutions and using a geometric series.

\section{Proof that interpolating terms cancel under integrals\label{cancel}}

We want to show that the contribution in the Pad\'e approximation beyond the extended quasiparticle picture in (\ref{excess}) is negligible. This means we consider
\ba
\langle\phi\rangle&=\int{dk} \phi_k\rho^{\rm Pade}(k)=\langle\phi\rangle^{\rm EQP}
\nonumber\\&+\int dk \phi_k f\left [{\epsilon_F-\varepsilon_k\over {\gamma(\epsilon_F,k)\over 2}}\right ] \int {d\omega \over 2 \pi} n_\omega [a_{\rm f}(\omega,k)-a_{\rm 1-f}(\omega,k)].
\end{align}
Since the interpolating function has the value $f=1$ only in a small limited neighborhood of the Fermi energy $\varepsilon_k=\epsilon_F$ we can write the additional term
\ba
\langle\phi\rangle\!-\!\langle\phi\rangle^{\rm EQP}\!\approx\! \phi_{k_F}\!\!\int \!\!\!{d\omega \over 2 \pi} n_\omega [a_{\rm f}(\omega,k)\!-\!a_{\rm 1\!-\!f}(\omega,k)]_{\varepsilon_k=\epsilon_F}.
\label{add}
\end{align}
Similar to the steps leading to the frequency-weighted sum rule (\ref{sumw})
we make a partial integration observing $\varepsilon_k=\varepsilon_k^0+\sigma$ with (\ref{eps0}) to have
\ba
&\int {d \omega \over 2 \pi} n_\omega a_{\rm 1-f}(\omega,k)=n_{\varepsilon_k}-\sigma n_{\varepsilon_k}'\nonumber\\&
+\int{d\omega \over 2 \pi} {\gamma(\omega,k)(n_\omega'\!-\!n_{\varepsilon_k}')\!+\!\p\omega\gamma(\omega,k)(n_\omega\!-\!n_{\varepsilon_k})\over \omega \!-\!\varepsilon_k}\!+\!o(\gamma^2)
\label{a1}
\end{align}
where the prime denotes the derivative and abbreviating  $\gamma=\gamma(\epsilon_F,k)$
\ba
&\int {d \omega \over 2 \pi} n_\omega a_{\rm f}(\omega,k)=n_{\epsilon_F}+(\varepsilon_k^0-\epsilon_F) n_{\epsilon_F}'\nonumber\\&
+\!\!\int\!\!{d\omega \over 2 \pi} {\gamma\over (\omega \!-\!\epsilon_F)^2\!+\!{\gamma^2 \over 4}}\left [(n_\omega\!-\!n_{\epsilon_F})\!+\!(\varepsilon_k^0\!-\!\epsilon_F)(n_\omega'\!-\!n_{\epsilon_F}')\right ].
\label{a2}
\end{align}
In the last expression we consider it as a first-order Taylor expansion and write $[...]\approx n_{\omega+\varepsilon_k^0-\epsilon_F}-n_{\varepsilon_k^0}$. This expression is integrated with the Lorentzian weight in the second term of (\ref{a2}) which peaks at $\omega\approx\epsilon_F$ where the expression vanishes. Therefore the second term of (\ref{a2}) is zero up to orders of $o(\gamma(\epsilon_F)^2)$. 

Needed in (\ref{add}) we have to subtract from (\ref{a1}) at $\varepsilon_k=\epsilon_F$ the term (\ref{a2}). Since $\varepsilon_k=\varepsilon_k^0+\sigma$ it remains only
\be
&&\int \!\!{d\omega \over 2 \pi} n_\omega [a_{\rm f}(\omega,k)\!-\!a_{\rm 1\!-\!f}(\omega,k)]_{\varepsilon_k=\epsilon_F}
\nonumber\\&&
=
\int{d\omega \over 2 \pi} {\gamma(\omega,k)(n_\omega'-n_{\epsilon_F}')+\p\omega\gamma(\omega,k)(n_\omega-n_{\epsilon_F})\over \omega -\epsilon_F}.
\nonumber\\&&
=
\int{d\omega \over 2 \pi} {\gamma(\omega,k)\over (\omega -\epsilon_F)^2}[(n_\omega-n_{\epsilon_F})-n_{\epsilon_F}'(\omega-\epsilon_F)]
\ee
where we used a partial integration to go to the third line. Since the integral is peaked around $\omega\sim\epsilon_F$ we can expand the first term in the $[...]$ bracket up to first order which cancels exactly and we can conclude that (\ref{add}) is vanishing.

\section*{Data Availability Statement}
No Data associated in the manuscript.


\begin{thebibliography}{10}

\bibitem{SSH01}
T. Sch{\"a}tz, U. Schramm, and D. Habs, Nature {\bf 412},  717  (2001).

\bibitem{SSBH02}
U. Schramm, T. Sch{\"a}tz, M. Bussmann, and D. Habs, Plasma Phys. Control
  Fusion {\bf 44},  B375  (2002).

\bibitem{PMC14}
M. Pagano, G.and~Mancini, G. Cappellini, and et~al., Nature Phys {\bf 10},  198
   (2014).

\bibitem{Saito98}
R. Saito, G. Dresselhaus, and M.~S. Dresselhaus, {\em Physical Properties of
  Carbon Nanotubes} (Imperial College Press, London, 1998).

\bibitem{Bockrath99}
M. Bockrath {\it et~al.}, Nature {\bf {\bf397}},  598  (1999).

\bibitem{Ishii03}
H. Ishii {\it et~al.}, Nature {\bf {\bf 426}},  540  (2003).

\bibitem{Shiraishi03}
M. Shiraishi and M. Ata, Sol. State Commun. {\bf {\bf 127}},  215  (2003).

\bibitem{Milliken96}
F.~P. Milliken, C.~P. Umbach, and R.~A. Webb, Sol. State Commun. {\bf {\bf
  97}},  309  (1996).

\bibitem{Mandal01}
S.~S. Mandal and J.~K. Jain, Sol. State Commun. {\bf {\bf 118}},  503  (2001).

\bibitem{Chang03}
A.~M. Chang, Rev. Mod. Phys. {\bf {\bf 75}},  1449  (2003).

\bibitem{Schafer08}
J. Sch{\"{a}}fer {\it et~al.}, Phys. Rev. Lett. {\bf {\bf 101}},  236802
  (2008).

\bibitem{Huang01}
Y. Huang {\it et~al.}, Science {\bf {\bf 294}},  1313  (2001).

\bibitem{Monien98}
H. Monien, M. Linn, and N. Elstner, Phys. Rev. A {\bf {\bf 58}},  R3395
  (1998).

\bibitem{Recati03}
A. Recati, P.~O. Fedichev, W. Zwerger, and P. Zoller, J. Opt. B: Quantum
  Semiclass. Opt. {\bf {\bf 5}},  S55  (2003).

\bibitem{Moritz05}
H. Moritz {\it et~al.}, Phys. Rev. Lett. {\bf {\bf 94}},  210401  (2005).

\bibitem{Nitzan03}
A. Nitzan and M.~A. Ratner, Science {\bf {\bf 300}},  1384  (2003).

\bibitem{Lu77}
A. Luther, Phys. Rev. B {\bf 15},  403  (1977).

\bibitem{Solyom79}
J. S{\'o}lyom, Adv. Phys. {\bf 28},  209  (1979).

\bibitem{GGM10}
D.~B. Gutman, Y. Gefen, and A.~D. Mirlin, Phys. Rev. B {\bf 81},  085436
  (2010).

\bibitem{H81a}
F.~D.~M. Haldane, Phys. Rev. Lett. {\bf 47},  1840  (1981).

\bibitem{L63}
J.~M. Luttinger, J. Math. Phys. {\bf 4},  1154  (1963).

\bibitem{LP74}
A. Luther and I. Peschel, Phys. Rev. B {\bf 9},  2911  (1974).

\bibitem{ES07}
U. Eckern and P. Schwab, phys. stat. sol. (b) {\bf 244},  2343  (2007).

\bibitem{DL74}
I.~E. Dzyaloshinskii and A.~I. Larkin, Sov. Phys. JETP {\bf 38},  202  (1974).

\bibitem{Ram17}
L. Rammelm\"uller, W.~J. Porter, J. Braun, and J.~E. Drut, Phys. Rev. A {\bf
  96},  033635  (2017).

\bibitem{EFGKK10}
F.~H.~L. E\ss{}ler {\it et~al.}, {\em The One-dimensional Hubbard Model}
  (Cambridge University Press, Cambridge, 2010).

\bibitem{GBML13}
X.-W. Guan, M.~T. Batchelor, and C. Lee, Rev. Mod. Phys. {\bf 85},  1633
  (2013).

\bibitem{LD11}
R.~M. Lee and N.~D. Drummond, Phys. Rev. B {\bf 83},  245114  (2011).

\bibitem{LG16}
P.~F. Loos and M.~W. Gill, WIREs Comput. Mol. Sci. {\bf 6},  410  (2016).

\bibitem{Loos13}
P.-F. Loos, The Journal of Chemical Physics {\bf 138},  064108  (2013).

\bibitem{GADMP22}
A. Girdhar {\it et~al.}, Phys. Rev. B {\bf 105},  115140  (2022).

\bibitem{GPS08}
S. Giorgini, L.~P. Pitaevskii, and S. Stringari, Rev. Mod. Phys. {\bf 80},
  1215  (2008).

\bibitem{T67}
A. Theumann, J. Math. Phys. {\bf 8},  2460  (1967).

\bibitem{VMKA08}
V. Garg, R.~K. Moudgil, K. Kumar, and P.~K. Ahluwalia, Phys. Rev. B {\bf 78},
  045406  (2008).

\bibitem{Sch77}
P. Schlottmann, Phys. Rev. B {\bf 16},  2055  (1977).

\bibitem{Yo01}
K. Yokoyama, Journal of the Physical Society of Japan {\bf 70},  2825  (2001).

\bibitem{V94}
J. Voit, Rep. Prog. Phys. {\bf 57},  977  (1994).

\bibitem{G04}
T. Giamarchi, Chem. Rev. {\bf 104},  5037  (2004).

\bibitem{GV08}
G.~F. Giuliani and G. Vignale, {\em Quantum theory of electron liquid}
  (Cambridge University Press, Cambridge, 2008).

\bibitem{C66a}
R.~A. Craig, Ann. Phys. {\bf 40},  416  (1966).

\bibitem{SZ79}
H. Stolz and R. Zimmermann, phys. stat. sol. (b) {\bf 94},  135  (1979).

\bibitem{KKL84}
D. Kremp, W.~D. Kraeft, and A.~D.~J. Lambert, Physica A {\bf 127},  72  (1984).

\bibitem{SR87}
M. Schmidt and G. R{\"o}pke, phys. stat. sol. (b) {\bf 139},  441  (1987).

\bibitem{MR94}
K. Morawetz and G. Roepke, Phys. Rev. E {\bf 51},  4246  (1995).

\bibitem{BKKS96}
T. Bornath, D. Kremp, W.~D. Kraeft, and M. Schlanges, Phys. Rev. E {\bf 54},
  3274  (1996).

\bibitem{SLM96}
V. {\v S}pi{\v c}ka, P. Lipavsk{\'y}, and K. Morawetz, Phys. Lett. A {\bf 240},
   160  (1998).

\bibitem{MLS00}
K. Morawetz, P. Lipavsk{\'y}, and V. {\v S}pi{\v c}ka, Ann. of Phys. {\bf 294},
   135  (2001).

\bibitem{M17b}
K. Morawetz, {\em Interacting systems far from equilibrium - quantum kinetic
  theory} (Oxford University Press, Oxford, 2017).

\bibitem{SLM98}
V. {\v S}pi{\v c}ka, K. Morawetz, and P. Lipavsk{\'y}, Phys. Rev. E {\bf 64},
  046107  (2001).

\bibitem{MLSK98}
K. Morawetz, P. Lipavsk{\'y}, V. {\v S}pi{\v c}ka, and N.~H. Kwong, Phys. Rev.
  C {\bf 59},  3052  (1999).

\bibitem{M17}
K. Morawetz, Phys. Rev. E {\bf 96},  032106  (2017).

\bibitem{SDC19}
S. Liang, D. Zhang, and W. Chen, J. Phys. Condens. Matter {\bf 31},  185601
  (2019).

\bibitem{DBG10}
V. Deshpande, M. Bockrath, L. Glazman, and et~al., Nature {\bf 464},  209
  (2010).

\bibitem{GVM93}
J. Gal\'an, J.~A. Verg\'es, and A. Martin-Rodero, Phys. Rev. B {\bf 48},  13654
   (1993).

\bibitem{FA22}
M. Fabrizio, Nat Commun {\bf 13},  1561  (2022).

\bibitem{Schu90}
H.~J. Schulz, Phys. Rev. Lett. {\bf 64},  2831  (1990).

\bibitem{BVB90}
M. Brech, J. Voit, and H. B{\"u}ttner, Europhys. Lett. {\bf 12},  289  (1990).

\bibitem{BS19}
D. Buterakos and S. Das~Sarma, Phys. Rev. B {\bf 100},  235149  (2019).

\bibitem{GS67}
H. Gutfreund and P. Schick, Phys. Rev. {\bf 168},  418  (1967).

\bibitem{OD03}
M. Olshanii and V. Dunjko, Phys. Rev. Lett. {\bf 91},  090401  (2003).

\bibitem{M02}
S.~A. Morgan, M.~D. Lee, and K. Burnett, Phys. Rev. A {\bf 65},  022706
  (2002).

\bibitem{SZ19}
Y. Song and H. Zhang, Eur. Phys. J. D {\bf 73},  106  (2019).

\bibitem{Fulde64}
P. Fulde and R.~A. Ferrel, Phys. Rev. {\bf 135},  A550  (1964).

\bibitem{Larkin65}
A.~I. Larkin and Y.~N. Ovchinnikov, Sov. Phys. JETP {\bf 20},  762  (1965).

\bibitem{PAC16}
A. Petrovi{\'c}, D. Ansermet, D. Chernyshov, and et~al., Nat Commun {\bf 7},
  12262  (2016).

\bibitem{Be11}
B. Bergk {\it et~al.}, New Journal of Physics {\bf 13},  103018  (2011).

\bibitem{WSLS12}
Z. Wang, W. Shi, R. Lortz, and P. Sheng, Nanoscale {\bf 4},  21  (2012).

\bibitem{He15}
M. He {\it et~al.}, Journal of Physics: Condensed Matter {\bf 27},  075702
  (2015).

\bibitem{PHC16}
D. Ansermet {\it et~al.}, ACS Nano {\bf 10},  515  (2016).

\bibitem{LW72}
D.~C. Langreth and J.~W. Wilkins, Phys. Rev. B {\bf 6},  3189  (1972).

\bibitem{Fa21}
B. Farid, 2021.

\bibitem{SL94}
V. {\v S}pi{\v c}ka and P. Lipavsk{\'y}, Phys. Rev. Lett. {\bf 73},  3439
  (1994).

\bibitem{SL95}
V. {\v S}pi{\v c}ka and P. Lipavsk{\'y}, Phys. Rev. B {\bf 52},  14615  (1995).

\bibitem{SLMa96}
V. {\v S}pi{\v c}ka, P. Lipavsk{\'y}, and K. Morawetz, Phys. Rev. B {\bf 55},
  5084  (1997).

\bibitem{SLMb96}
V. {\v S}pi{\v c}ka, P. Lipavsk{\'y}, and K. Morawetz, Phys. Rev. B {\bf 55},
  5095  (1997).

\bibitem{Fa19}
B. Farid, 2019.

\bibitem{HuS93}
B.~Y.-K. Hu and S. Das~Sarma, Phys. Rev. B {\bf 48},  14388  (1993).

\bibitem{M68}
N.~F. Mott, Philosph. Mag. {\bf 19},  835  (1969).

\bibitem{RM21}
Z. Ristivojevic and K.~A. Matveev, Phys. Rev. Lett. {\bf 127},  086803  (2021).

\bibitem{Mal78}
P.~F. Maldaque, Surface Science {\bf 73},  296  (1978).

\end{thebibliography}

\end{document}